\documentclass{elsart}
\pdfoutput=1
\usepackage{natbib}
\usepackage{amssymb}
\usepackage{amsmath}
\usepackage{graphicx}
\usepackage[center,small]{subfigure}
\usepackage{color}
\usepackage{hyperref}
\newcommand{\ds}{\displaystyle}
\begin{document}
\begin{frontmatter}

\title{Configurational stability of a crack propagating in a material with mode-dependent fracture energy - Part II: Drift of fracture facets in mixed-mode I+II+III}

\author{Aditya Vasudevan\corauthref{cor}$^{1,2}$},
\author{Laurent Ponson$^1$},
\author{Alain Karma$^2$},
\author{Jean-Baptiste Leblond$^1$}
\corauth[cor]{Corresponding author.}

\address{$^1$Sorbonne Universit\'{e}, Facult\'{e} des Sciences et Ing\'{e}nierie, Campus Pierre et Marie Curie, CNRS, UMR 7190, Institut Jean Le Rond d'Alembert, F-75005 Paris, France}
\address{$^2$Physics Department and Center for Interdisciplinary Research on Complex Systems, Northeastern University, Boston, MA 02115, USA}

\begin{abstract}

In earlier papers \citep{LKL11,LKPV18}, we presented linear stability analyses of the coplanar propagation of a crack loaded in mixed-mode I+III, based on a ``double'' propagation criterion combining \cite{G20}'s energetic condition and \cite{GS74}'s principle of local symmetry. The difference between the two papers was that in the more recent one, the local value of the critical energy-release-rate was no longer considered as a constant, but heuristically allowed to depend upon the ratio of the local mode III to mode I stress intensity factors. This led to a much improved, qualitatively acceptable agreement of theory and experiments, for the ``threshold'' value of the ratio of the unperturbed mode III to mode I stress intensity factors, above which coplanar propagation becomes unstable. In this paper, the analysis is extended to the case where a small additional mode II loading component is present in the initially planar configuration of the crack, generating a small, general kink of this crack from the moment it is applied. The main new effect resulting from presence of such a loading component is that the instability modes present above the threshold must drift along the crack front during its propagation. This prediction may be useful for future theoretical interpretations of a number of experiments where such a drifting motion was indeed observed.

\end{abstract}

\end{frontmatter}


\section{Introduction}\label{sec:Intro}

Fracture of materials leads to a myriad of different patterns including oscillatory crack paths~\citep{YS93,LBF07}, self-similar fracture surfaces~\citep{BLP90,SRMSS02}, star-shaped patterns~\citep{Vandenberghe} and periodic columnar structures~\citep{GMM09,GLP10}. The ensemble of patterns formed by fracture provides a readily available benchmark to assess and discriminate competing models of material failure. The study of these patterns has already resulted in major advances in the understanding of a wide variety of fracture problems, ranging from the stability of tensile cracks~\citep{YR01,CBHK09} and the role of nonlinear elasticity on crack growth~\citep{BLF08} to the effect of material heterogeneities on failure~\citep{Ponson16}.

Under mixed-mode conditions including an anti-plane shear component (mode I+III), a crack may fragment into an array of daughter cracks, leaving behind it a fracture surface that exhibits a factory roof profile. Since the seminal experimental work of~\cite{S69} on glass, such a pattern has been reported and studied in a wide variety of materials including metallic alloys~\citep{YM89,ERK17}, brittle polymers~\citep{CP96,LBFW08,LMR10} and rocks observed in situ~\citep{PSD82,NP85,Weinberger}, among others. A standard observation is the formation of a periodic array of facets parallel to each other but twisted around the direction of propagation towards a plane perpendicular to the principal stress axis. The inclination of the facets with respect to the mean fracture plane results in a significant reduction of the local mode III component in comparison to its global counterpart.

Recently, numerical simulations based on phase field methods have led to a more complete characterization of the geometry of segmented cracks. \cite{PK10}'s numerical simulations, based on \cite{KKL01}'s phase-field model that reproduces standard crack propagation laws of Linear Elastic Fracture Mechanics (LEFM) \citep{HK09}, revealed that a crack loaded in mode I+III gradually evolves from a planar configuration to a fragmented one, through unstable growth of helical perturbations. \cite{CCLNPK15} and \cite{PR14} provided a detailed picture of the coarsening mechanism that leads to the merging of adjacent facets into larger ones as the crack propagates. Yet, several features of the fragmentation patterns remain poorly understood. Such features notably include the distance between neighboring facets at the onset of fragmentation, the shape of the facets (ratio of their width to their length), the width of the ligaments connecting adjacent facets, and the geometrical (surfacic versus volumic) nature and extent of the damage within these ligaments.

Another important aspect of crack fragmentation under anti-plane shear is the ``critical'' or ``threshold'' value $\rho^\mathrm{cr}$ of the ``mixity ratio'' $\rho^0$, that is the ratio ${K_{III}^0}/{K_{I}^0}$ of the unperturbed mode III to mode I stress intensity factors (SIFs), above which facets start to form. Experimental studies report threshold values that strongly depend on the type of material, as $\rho^\mathrm{cr}$ does not exceed a few percent in glass~\citep{S69}, PMMA~\citep{PR14,Liu} and Homalite-100~\citep{LMR10}, but can be as large as $0.4$ in some aluminum alloy~\citep{ERK17}.

The theoretical prediction of the fragmentation threshold and the resulting fracture pattern is a challenging task. An appealing approach, within the classical framework of LEFM, consists in performing a linear stability analysis of a crack propagating under mixed-mode I+III conditions. In this approach, one looks for crack front configurations satisfying, at all points of the front and all instants during propagation, both \cite{G20}'s energetic condition and \cite{GS74}'s Principle of Local Symmetry (PLS) stipulating that the local mode II SIF must vanish. Following this line of thought, \cite{LKL11} showed that the initially straight configuration of the crack front becomes unstable above some critical mode mixity ratio $\rho^\mathrm{cr}$, and then bifurcates into a helical geometry with an exponentially growing amplitude. The value of $\rho^\mathrm{cr}$ predicted depends only on that of Poisson's ratio $\nu$. The prediction of bifurcated modes also sheds some light on the geometry of fragmented crack fronts: indeed helical perturbations of small wavelength are found to be the least stable of all, implying that fragmentation must initiate
at a small lengthscale that is set by the size of the process zone where fracture occurs \citep{Barenblatt} - consistent with both experimental observations~\citep{PR14,ERK17} and numerical simulations~\citep{PK10}. The latter also indicated that linear perturbations of the crack front become stable below a critical wavelength that scales proportionally to the process zone size but also generally depends on mode mixity \citep{PK10}.

However, the comparison of the predicted threshold and that actually observed is less successful, as the theory largely overestimates, in most materials, the amount of mode mixity required to fragment the crack, predictions being in the range $\rho^\mathrm{cr} \simeq 0.4-0.5$ for $\nu \simeq 0.3-0.4$. Also, the rather modest variations of Poisson's ratio from one material to another do not seem capable of explaining the wide variations of the threshold actually observed; a clear indication that this threshold may in reality depend on additional material parameters.

A possible interpretation of the discrepancy between theoretical and experimental values of the threshold was recently provided by \cite{CCLNPK15}, who extended \cite{PK10}'s simulations based on \cite{KKL01}'s phase-field model by performing an extensive study of non-coplanar solutions. In this work the bifurcation accompanying the transition from coplanar to fragmented front was shown to be strongly subcritical, which suggested that jumps from the stable branch to the unstable one could be induced well below the theoretical threshold by large enough perturbations.

Even more recently, following a different line of thought, \cite{LKPV18} revisited \cite{LKL11}'s stability analysis by accounting for a new physical mechanism: instead of assuming the fracture energy to be a constant, they introduced a possible dependence of this energy upon the local mixity ratio. This new heuristic hypothesis was motivated indirectly by \cite{Freund2}'s and \cite{FGBP17}'s observation that interfacial\footnote{Propagation of the crack along an interface warrants that it does not kink so as to eliminate mode II.} fracture energy significantly increases with the amount of plane shear applied (mode II)~, and more straightforwardly by the toughening observed in the presence of anti-plane shear (mode III) by \cite{Liu} and  \cite{DS93} , for PMMA, \cite{LMR10}, for Homalite, and \cite{Suresh}, for alumina. A material parameter $\gamma$ characterizing the toughening induced by anti-plane shear was thus introduced. It was found that in the presence of mode III, a shear-dependent fracture energy, by making crack propagation more difficult along a plane and easier along facets with a reduced mode III SIF, results in an earlier formation of tilted facets. This effect may reconcile predicted and observed values of the fragmentation threshold and explain in particular, {\it via} its additional dependence upon the parameter $\gamma$, its wide variations from one material to another \citep{LKPV18}.

\begin{figure}[h]
\centerline{\includegraphics[height=10cm]{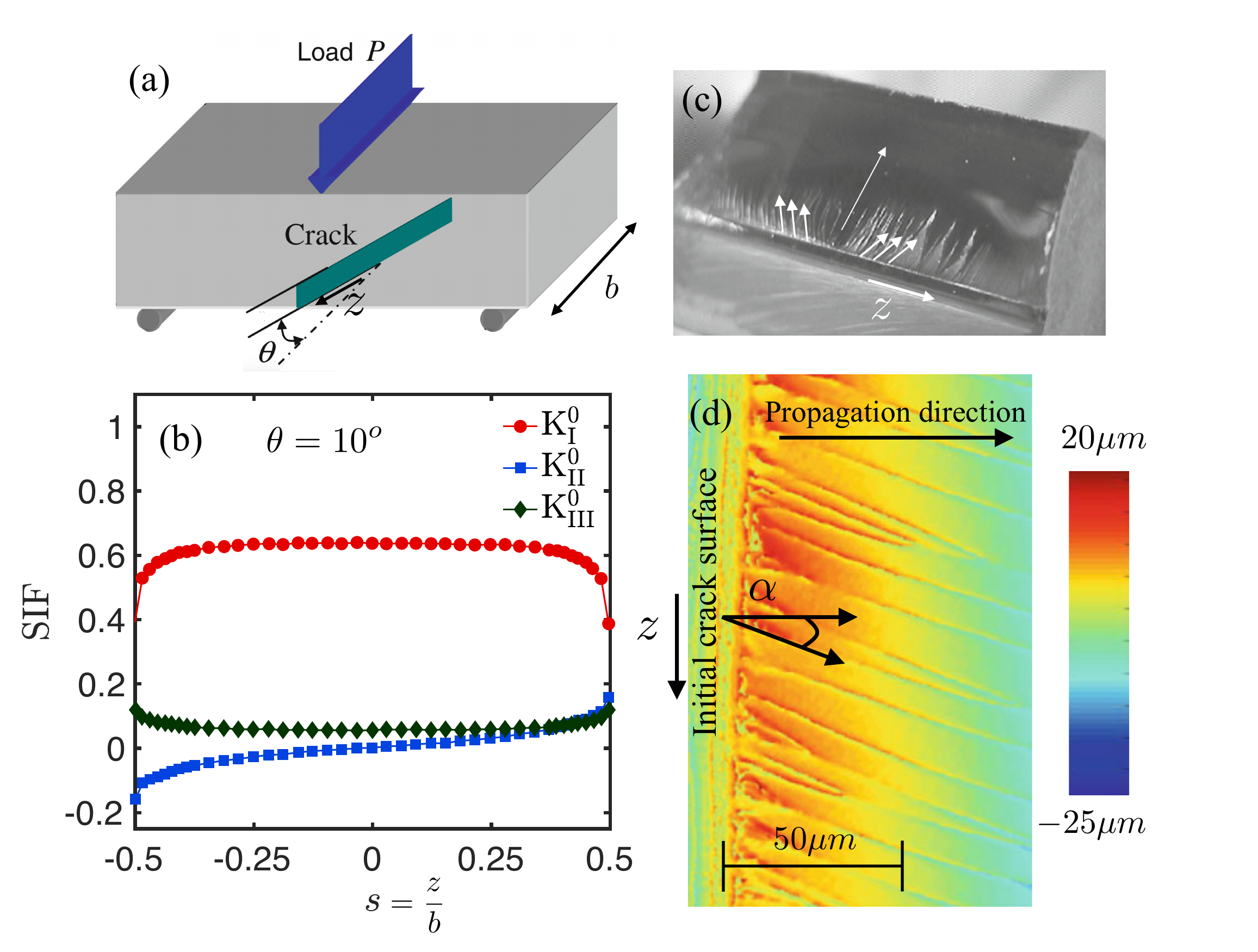}} \caption{Experimental setup of a three point bending mixed-mode test on Homalite (Courtesy of \cite{LMR10}). (a) shows the schematic with a tilted crack at an angle $ \theta $ while (b) shows the distribution of the stress intensity factors (SIF) in mode I, II and III for $ \theta = 10^o$ along the crack front normalized by the width of the sample. (c) and (d) show the post-mortem fracture surface pattern of the facets along the whole width and in a local region with non-zero mode II, respectively, where facets are clearly observed to be drifting at a varying angle with respect to the mean crack front propagation direction.}
\label{fig:ExpSetup}
\end{figure}

In this paper, we pursue the analysis beyond the study of the fragmentation threshold, by now concentrating on the fracture pattern predicted by the LEFM-based model. The focus is essentially on a generalization of the previous linear stability analysis to completely general (I+II+III) mixed-mode conditions. Such a generalization is also motivated by three point bending experiments of a tilted notch (see Fig. \ref{fig:ExpSetup}) where the tilted notch imposes mode III, but there is also a non-negligible amount of mode II induced that is zero at the center and varies gradually along the crack front (Fig. \ref{fig:ExpSetup}(b)). Fig. \ref{fig:ExpSetup}(c)  and Fig. \ref{fig:ExpSetup}(d) show the resulting fracture surface pattern. In addition to facet formation, this pattern reveals that facets drift at an angle $ \alpha $ from the propagation direction. In this work, through linear stability calculations, we revisit the analysis of \cite{LKPV18} under general (I+II+III) mixed-mode conditions. The results of this analysis indeed predict that, in the presence of mode II, the facets drift along the front as the crack propagates, leaving behind ridges that are no longer parallel to the mean direction of crack propagation. We further explore the dependence of the drift angle (between the mean direction of crack propagation and the direction of the ridges) upon the amount of mode II, and also upon the parameter $\gamma$. Special attention is paid to the absence or presence of such a drifting motion in the absence of mode III-induced toughening ($\gamma = 0$). The results obtained is qualitatively compared with experiments of \cite{LMR10} and in the future  may provide theoretical grounds for the interpretation of various fragmentation patterns reported in the literature under mode I+II+III conditions~\citep{Sherman,LBFW08,Ronsin}, which show ridges obliquely oriented with respect to the mean direction of crack propagation. They also act as a motivation to perform new phase-field-based numerical simulations, including mode III-induced toughening and/or presence of a small mode II loading component. 

The paper is organized as follows:
\begin{itemize}
\item In Section~\ref{sec:PertCrack}, we define general hypotheses, and introduce first-order perturbation formulae for the SIFs along the front of a semi-infinite crack slightly but otherwise arbitrarily perturbed both within and out of its plane. These formulae are essentially adapted from the works of~\cite{GR86} for the in-plane perturbation and~\cite{MGW98} for the out-of-plane perturbation.
\item In Section~\ref{sec:FoundStabAnal}, we set up the foundations of our new linear stability analysis, extending that of~\cite{LKPV18} through incorporation of some additional mode II loading component.
\item We then present two distinct stability calculations: first, in Section~\ref{sec:StabAnal1} for a mode III-dependent fracture energy but a small mode III loading component; second, in Section~\ref{sec:StabAnal2} for a constant fracture energy but an arbitrary mode III component. We determine in particular, in both cases, the geometrical features of the unstable modes, and the value of the drift angle as a function of the various material and mechanical parameters.
\item The implications of our results are finally discussed in Section~\ref{sec:Discuss}. A preliminary, essentially qualitative comparison of theoretical predictions limited to the linear regime of instability, and experimental observations in the strongly nonlinear regime of well-developed facets, is provided.

\end{itemize}



\section{First-order perturbation of a semi-infinite crack in an infinite body}\label{sec:PertCrack}

General hypotheses, notations and basic formulae for the SIFs along the front of a slightly perturbed semi-infinite crack have been presented in full detail in Part I \citep{LKPV18}. A summarized presentation is provided here for completeness.

\begin{figure}[h]
\centerline{\includegraphics[height=8 cm]{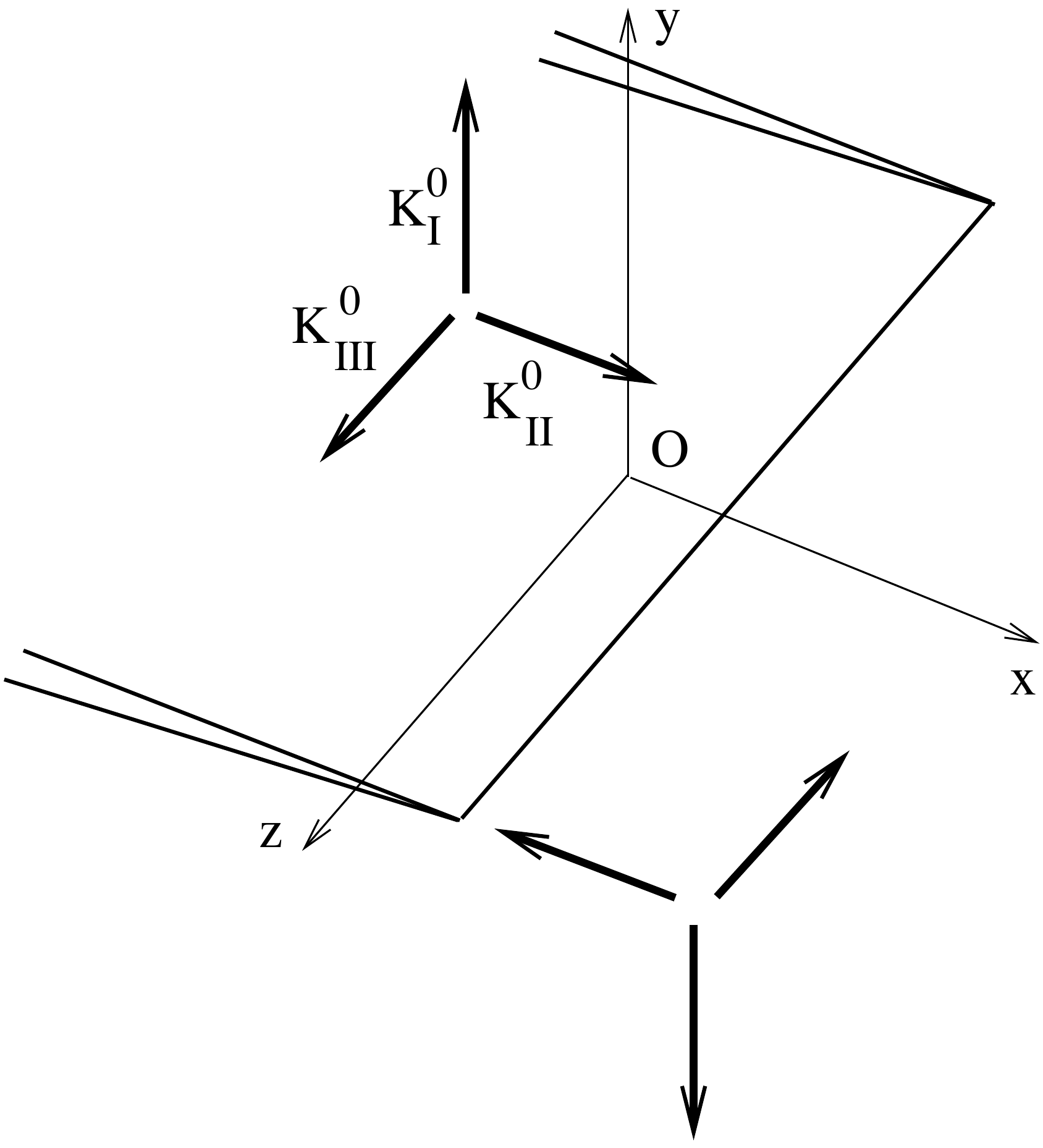}} \caption{Unperturbed geometry and loading.}
\label{fig:InitConfig}
\end{figure}

A semi-infinite crack embedded within an infinite isotropic elastic body is considered in two configurations. In the first, unperturbed one, the crack is planar and its front is straight (Fig. \ref{fig:InitConfig}). A Cartesian frame $(Oxyz)$ with axes oriented according to the standard convention is introduced. The crack is loaded under general mixed-mode I+II+III conditions, with uniform SIFs $K_I^0$, $K_{II}^0$, $K_{III}^0$ along the front.

\begin{figure}[h]
\centering
\subfigure[\label{fig:InPlanePert} In-plane perturbation of the crack front.]
{
\includegraphics[height=8cm]{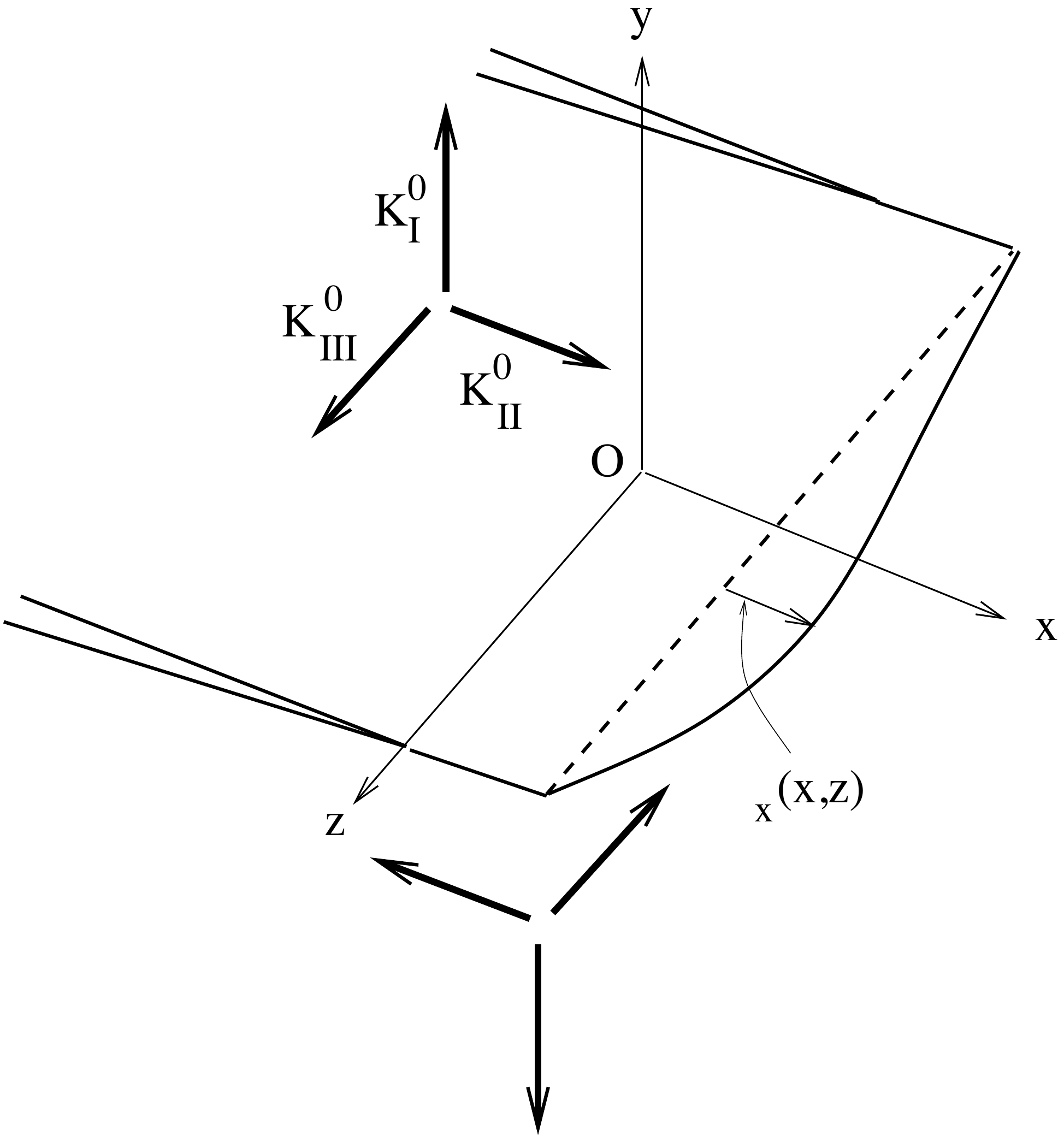}
}
\subfigure[\label{fig:OutOfPlanePert} Out-of-plane perturbation of the crack surface.]
{
\includegraphics[height=8cm]{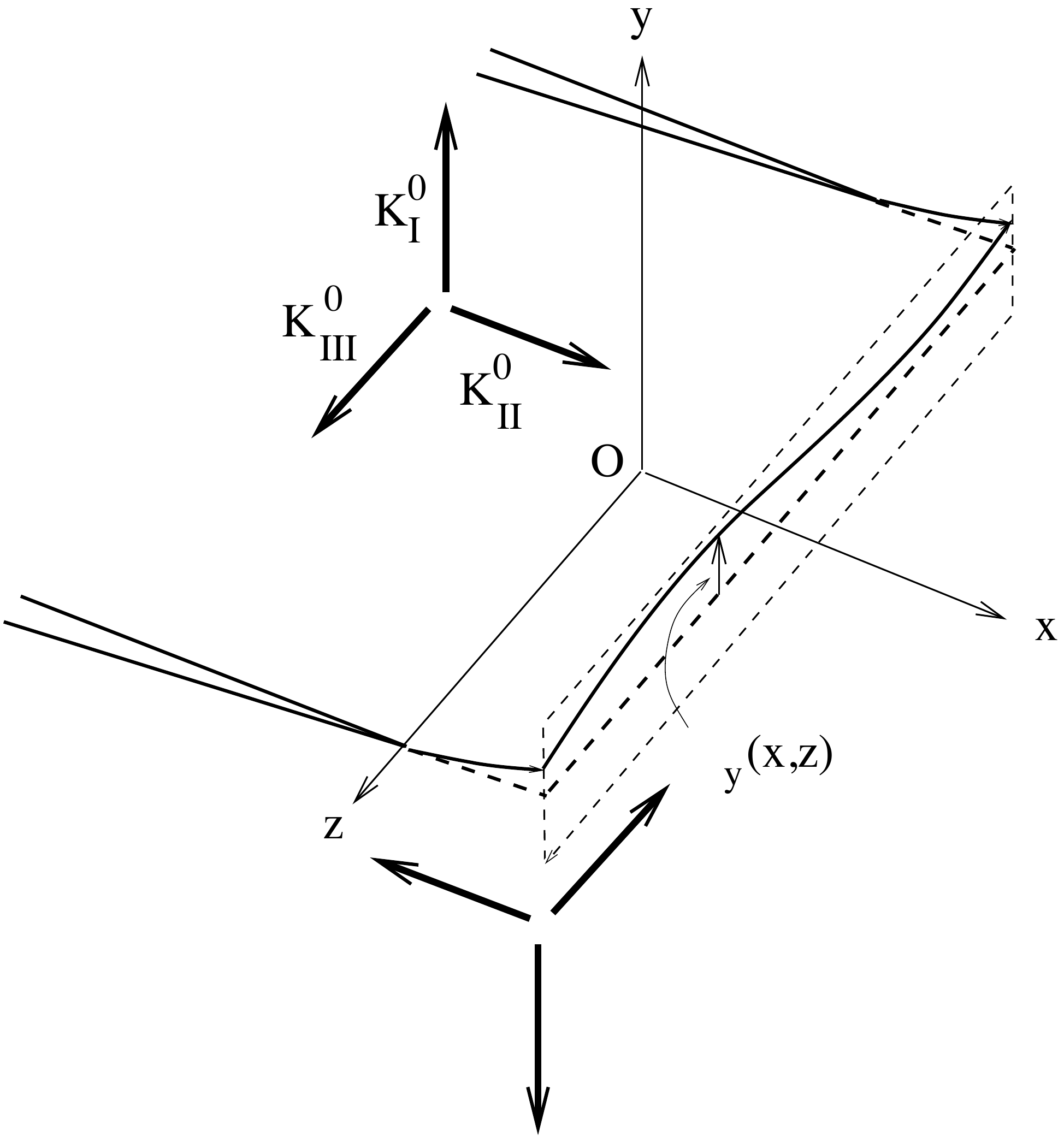}
}
\caption{\label{fig:PertConfig} In-plane and out-of-plane perturbations of the crack front and surface.}
\end{figure}

In the second, slightly perturbed configuration, the front of the crack is displaced in the direction $x$ by a small distance $\phi_x(x,z)$ (Fig. \ref{fig:InPlanePert}), and its surface is displaced in the direction $y$ by a small distance $\phi_y(x,z)$ (Fig. \ref{fig:OutOfPlanePert}).

In this new configuration, the perturbation $\delta K_{p}(x,z)$ of the $p$-th SIF is given, to first order in the pair $(\phi_x,\,\phi_y)$, by the formula
\begin{equation}\label{eqn:AddPertSIF}
\delta K_{p}(x,z) = \delta_{x}K_{p}(x,z) + \delta_{y}K_{p}(x,z) \quad (p=I,II,III)
\end{equation}
where the contributions $\delta_{x}K_{p}(x,z)$ and $\delta_{y}K_{p}(x,z)$ arise from $\phi_x$ and $\phi_y$, respectively. With some mildly restrictive hypotheses detailed in \citep{LKPV18}, the contributions $\delta_{x}K_{p}(x,z)$ ($p=I,II,III$) due to $\phi_x$ are given by \cite{GR86}'s formulae:
\begin{equation}\label{eqn:dxK}
\left\{
\begin{array}{lll}
\ds \delta_xK_{I}(x,z) & = & \ds \frac{K_I^0}{2\pi} \, PV \int_{-\infty}^{+\infty}\frac{\phi_x(x,z')-\phi_x(x,z)}{(z'-z)^2}\,dz' \\[3.mm]
\ds \delta_xK_{II}(x,z) & = & \ds - \frac{2}{2-\nu} K_{III}^0\,\frac{\partial \phi_x}{\partial z}(x,z)
+ \frac{2-3\nu}{2-\nu}\,\frac{K_{II}^0}{2\pi} \, PV \int_{-\infty}^{+\infty}\frac{\phi_x(x,z')-\phi_x(x,z)}{(z'-z)^2}\,dz' \\[3.mm]
\ds \delta_xK_{III}(x,z) & = & \ds \frac{2(1-\nu)}{2-\nu} K_{II}^0\frac{\partial \phi_x}{\partial z}(x,z) + \frac{2+\nu}{2-\nu}\,\frac{K_{III}^0}{2\pi} \, PV \int_{-\infty}^{+\infty}\frac{\phi_x(x,z')-\phi_x(x,z)}{(z'-z)^2}\,dz' \\
\end{array}
\right.
\end{equation}
where $\nu$ denotes Poisson's ratio and the symbol $PV$ a Cauchy Principal Value. Also, the contributions $\delta_{y}K_{p}(x,z)$ ($p=I,II,III$) due to $\phi_y$ are given by \cite{MGW98}'s formulae:
\begin{equation}\label{eqn:dyK}
\left\{
\begin{array}{lll}
\ds \delta_yK_I(x,z) & = & \ds - \frac{3}{2}K_{II}^0\frac{\partial \phi_y}{\partial x}(x,z) - 2K_{III}^0\,\frac{\partial \phi_y}{\partial z}(x,z) \\[3.mm]
{} & {} & \ds - \frac{K_{II}^0}{2\pi} \, PV \int_{-\infty}^{+\infty}\frac{\phi_y(x,z')-\phi_y(x,z)}{(z'-z)^2}\,dz'
+ \delta_y K_I^{\rm skew}(x,z) \\[3.mm]
\ds \delta_yK_{II}(x,z) & = & \ds \frac{K_{I}^0}{2}\,\frac{\partial \phi_y}{\partial x}(x,z)
- \frac{2-3\nu}{2-\nu}\frac{K_I^0}{2\pi} \, PV \int_{-\infty}^{+\infty}\frac{\phi_y(x,z')-\phi_y(x,z)}{(z'-z)^2}\,dz' \\[3.mm]
\ds \delta_yK_{III}(x,z) & = & \ds \frac{2(1-\nu)^2}{2-\nu}K_{I}^0\,\frac{\partial \phi_y}{\partial z}(x,z). \\
\end{array}
\right.
\end{equation}
The quantity $\delta_y K_I^{\rm skew}(x,z)$ here (connected to \cite{B87}'s {\it skew-symmetric} crack-face weight functions, whence the notation) is given by
\begin{equation}\label{eqn:dyK1skew}
\delta_y K_I^{\rm skew}(x,z) = \frac{\sqrt{2}}{4\pi} \, \frac{1-2\nu}{1-\nu} \, {\rm Re} \left\{\int_{-\infty}^{x} dx' \int_{-\infty}^{+\infty}
\frac{[K_{III}^0-i(1-\nu)K_{II}^0](\partial \phi_y/\partial z)(x',z')}{(x-x')^{1/2}\left[x-x'+i(z-z')\right]^{3/2}} \, dz' \right\}
\end{equation}
where the cut of the complex power function is along the half-line of negative real numbers \citep{MGW98,LKL11}. Note that in contrast to $\delta_{x}K_{p}(x,z)$ and $\delta_{y}K_{p}(x,z)$ $\quad (p=I,II,III)$, which depend only on the crack front shape, $\delta_y K_I^{\rm skew}(x,z)$ depends on the shape of the entire fracture surface behind the crack front.


\section{Foundations of the linear stability analysis}\label{sec:FoundStabAnal}

\subsection{Hypotheses on the propagation criterion}\label{subsec:HypCrit}

Just like in our previous works \citep{LKL11,LKPV18}, the prediction of the successive configurations of the crack, resulting from its mixed-mode propagation, will be based on a ``double'' criterion enforced all along the crack front and at all instants during propagation, consisting of:
\begin{itemize}
\item \cite{G20}'s condition $G(x,z)=G_\mathrm{c}(x,z)$ where $G(x,z)$ denotes the local energy-release-rate and $G_\mathrm{c}(x,z)$ its local ``critical'' value inducing propagation of the front;
\item \cite{GS74}'s PLS according to which the local SIF $K_{II}(x,z)$ of mode II must be zero.
\end{itemize}

Like in our recent work \citep{LKPV18}, the critical energy-release-rate $G_\mathrm{c}$ will be allowed to possibly depend upon the ratio of the local mode III to mode I SIFs:\footnote{Note that it would make no sense to allow for an analogous dependence upon the ratio of the mode II to mode I SIFs, since the former SIF is necessarily zero by the PLS.}
\begin{equation}\label{eqn:DefRho}
G_\mathrm{c}(x,z) \equiv G_\mathrm{c}[\rho(x,z)] \quad , \quad \rho(x,z) \equiv \frac{K_{III}(x,z)}{K_{I}(x,z)}
\end{equation}
according to the heuristic formula
\begin{equation}\label{eqn:GcVsRho}
G_\mathrm{c}(\rho) \equiv G_\mathrm{Ic}(1 + \gamma |\rho|^{\kappa})
\end{equation}
where $G_\mathrm{Ic}$ denotes the value of $G_\mathrm{c}$ in pure mode I, and $\gamma$ and $\kappa$ positive, dimensionless material parameters. (The inequality $\gamma>0$ means that presence of mode III {\it increases} the value of $G_\mathrm{c}$).

\subsection{Hypotheses on the geometry and loading}\label{subsec:HypGeomLoad}

The general mixed-mode I+II+III conditions (in the planar configuration of the crack) considered in this paper will make the situation somewhat different from, and more complex than, that resulting from the mode I+III conditions considered in earlier papers \citep{LKL11,LKPV18}. This is due to the general kink of the crack induced by the presence of mode II, which will be considered to occur only once the crack front has reached a certain specific position.

\begin{figure}[h]
\centering
\subfigure[\label{fig:BeforeProp} Before propagation.]
{
\includegraphics[height=5cm]{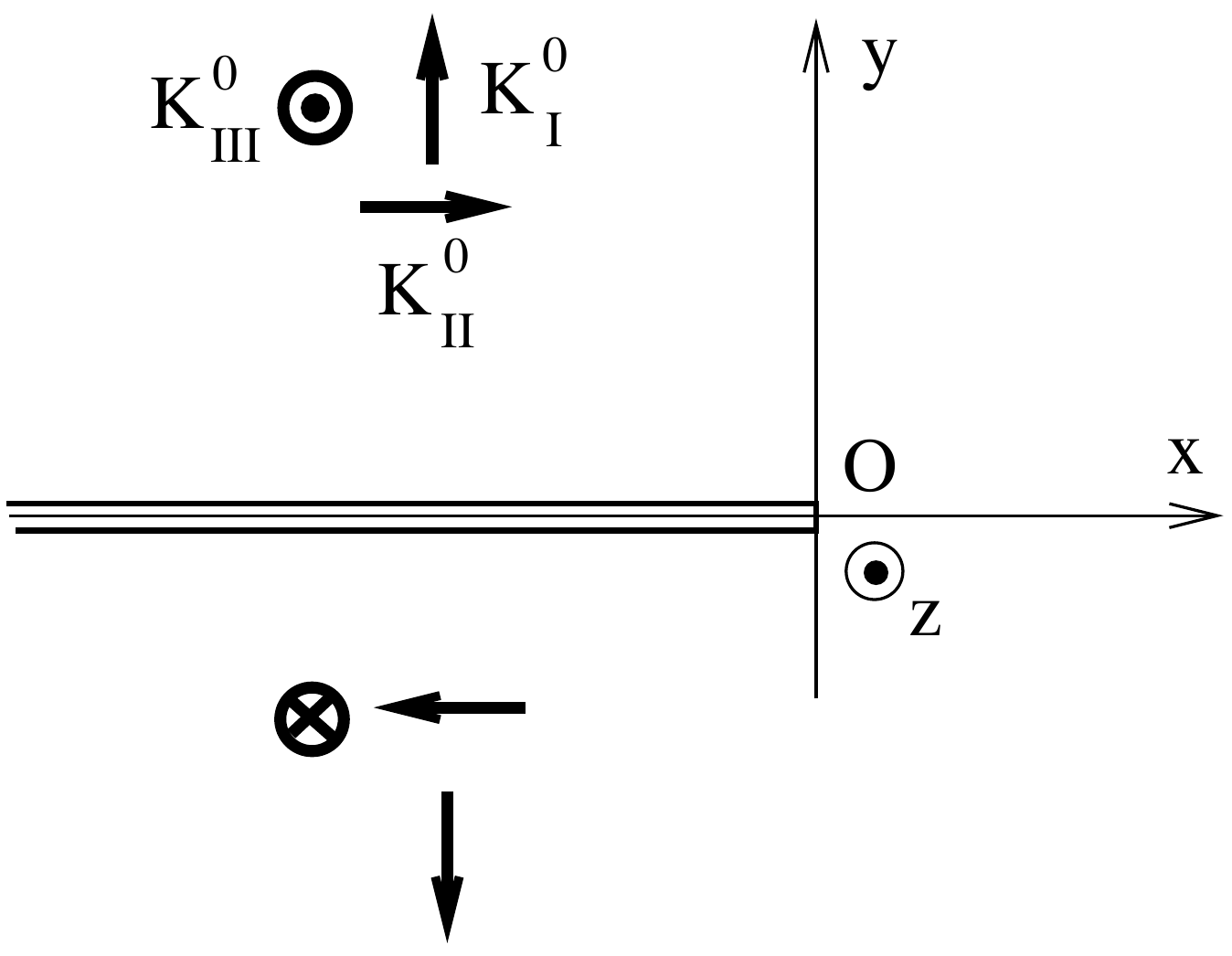}
}
\subfigure[\label{fig:DuringProp} During propagation.]
{
\includegraphics[height=5cm]{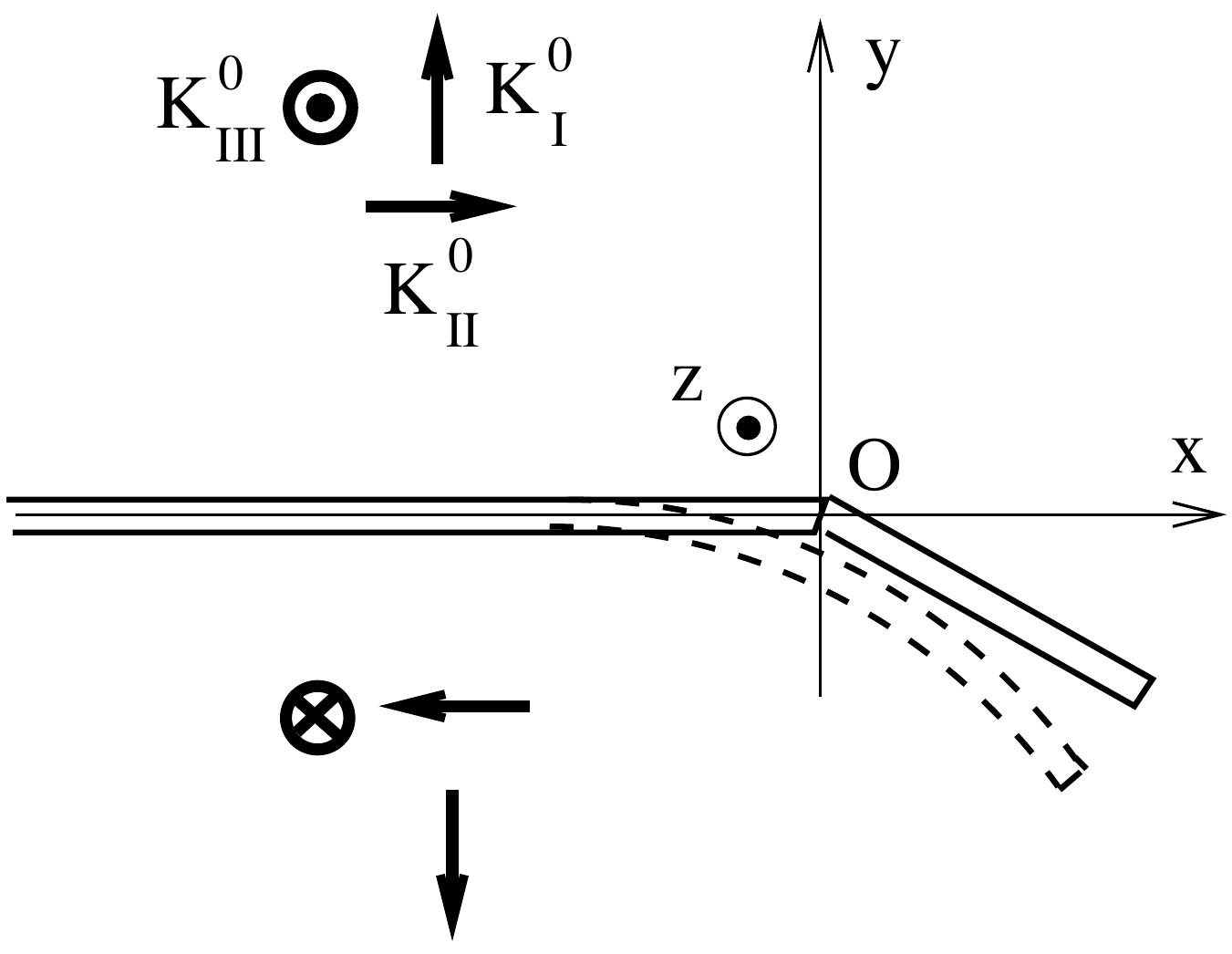}
}
\caption{\label{fig:BeforeAndDuringProp} Configurations of the crack before and during propagation in mixed-mode I+II+III.}
\end{figure}

More precisely, we shall consider an initially flat semi-infinite crack, occupying the domain $x<0$ within the plane $y=0$, obtained for instance through machining of the specimen or propagation in mode I fatigue. A static load including mode II and III components, of sufficient magnitude to induce crack propagation, will be assumed to be applied henceforward. A general kink of the crack will ensue, with possibly superimposed perturbations of the crack front and surface growing in an unstable manner. Figures \ref{fig:BeforeProp} and \ref{fig:DuringProp} provide 2D schematic illustrations, in the plane $Oxy$, of the configurations of the crack in its initial state and after some propagation under such conditions. In Figure \ref{fig:DuringProp} the full line represents the fundamental, kinked but unperturbed configuration, and the dotted line a kinked {\it and} perturbed configuration. Note that since the perturbation is assumed to already be nonzero at $x=0$, it must necessarily extend in the region $x<0$.

A remark pertaining to terminology is in order here. The words ``unperturbed'' and ``perturbed'' have just been used in reference to the ``perturbation'' of the crack from its fundamental, {\it already kinked} configuration; this is logical in the context of a stability analysis devoted to the study of the growth or decay of the deviation of the crack from this configuration. However in \cite{MGW98}'s and \cite{LKL11}'s formulae (\ref{eqn:dyK}), (\ref{eqn:dyK1skew}), the out-of-plane ``perturbation'' to be considered {\it must include the additional contribution of the general kink}, since the reference crack in these formulae is strictly planar. It would be difficult to designate these two types of perturbation with distinct words; hence the same wording ``perturbation'' will be used in the sequel, the context making clear what is meant.

Following the notations introduced in the preceding Section, we denote $K_{I}^0$, $K_{II}^0$, $K_{III}^0$ the SIFs in the initial planar configuration of the crack (prior to mixed-mode propagation). These SIFs are assumed to be independent of the position $x$ of the crack front within the original crack plane - so that {\it if} the crack propagated along its original plane, the SIFs would retain their initial values $K_{I}^0$, $K_{II}^0$, $K_{III}^0$ at every instant. Without restricting generality, $K_{III}^0$ {\it may be assumed to be positive} like $K_{I}^0$. We then define the following dimensionless ratios:
\begin{equation}\label{eqn:DefRho0Phi0R0}
\varphi^0 \equiv \frac{K_{II}^0}{K_I^0} \quad ; \quad \rho^0 \equiv \frac{K_{III}^0}{K_I^0} \quad (>0) \quad ; \quad R^0 \equiv \frac{K_{II}^0}{K_{III}^0} \equiv \frac{\varphi^0}{\rho^0}.
\end{equation}
Note that the assumed positiveness of $\rho^0$ leaves the signs of $\varphi^0$ and $R^0$ arbitrary (though identical).

{\it The quantity $|\varphi^0|$ will be assumed to be much smaller than unity} - accordingly, terms of second order in $\varphi^0$ will be neglected in all formulae to follow. The reason, of technical nature, is tied to the fact that the mode II component of the loading generates a general kink angle proportional to $\varphi^0$ to first order. Thus if $|\varphi^0|$ were allowed to be large, the kink angle could also be large; and this would prohibit use of \cite{MGW98}'s and \cite{LKL11}'s first-order formulae (\ref{eqn:dyK}), (\ref{eqn:dyK1skew}) for the perturbed SIFs, which demand small ``slopes'' $\partial\phi_y/\partial x$, $\partial\phi_y/\partial z$ of the crack surface with respect to the initial crack plane $y=0$.


\subsection{Change of unknown function for the out-of-plane perturbation}\label{subsec:ChangeFunc}

The first task is to determine the general kink induced by the mode II loading component in the region $x\geq 0$ of propagation of the crack. By equation (\ref{eqn:dyK}), the local value of $K_{II}$ for a perturbation $\phi_y$ independent of $z$ is $K_{II} = K_{II}^0+\frac{K_I^0}{2}\frac{d\phi_y}{dx}$ so that by the PLS, the value of the kink angle (angle of rotation of the crack surface about the direction $z$ of the crack front) is $\frac{d\phi_y}{dx} = -2\frac{K_{II}^0}{K_{I}^0} = -2\varphi^0$. Thus the fundamental, kinked configuration of the crack consists of a semi-infinite crack occupying the half-plane $x<0$, $y=0$, supplemented in the region $x\geq 0$ with a kinked extension of equation $y=-2\varphi^0 x$ (Fig. \ref{fig:DuringProp}, full line).

To study deviations from this fundamental configuration, we introduce the change of unknown function for the out-of-plane perturbation defined by
\begin{equation}\label{eqn:DefPhiyTilde}
\left\{
\begin{array}{llll}
\phi_y(x,z) & = & \widetilde{\phi_y}(x,z) & \quad {\rm for}\ x<0 \\
\phi_y(x,z) & = & -2\varphi^0 x + \widetilde{\phi_y}(x,z) & \quad {\rm for}\ x\geq 0.
\end{array}
\right.
\end{equation}
With this new definition, \cite{MGW98}'s and \cite{LKL11}'s formulae (\ref{eqn:dyK}), (\ref{eqn:dyK1skew}) for the contribution of $\phi_y$ to the perturbations of the SIFs become in the region $x\geq 0$ (discarding in $\delta_yK_I$ a term proportional to $K_{II}^0\varphi^0$, of second order in $\varphi^0$):
\begin{equation}\label{eqn:dyK1}
\left\{
\begin{array}{lll}
\ds \delta_yK_I(x,z) & = & \ds - \frac{3}{2}K_{II}^0\frac{\partial \widetilde{\phi_y}}{\partial x}(x,z) - 2K_{III}^0\,\frac{\partial \widetilde{\phi_y}}{\partial z}(x,z) \\[3.mm]
{} & {} & \ds - \frac{K_{II}^0}{2\pi} \, PV \int_{-\infty}^{+\infty}\frac{\widetilde{\phi_y}(x,z')-\widetilde{\phi_y}(x,z)}{(z'-z)^2}\,dz'
+ \delta_y K_I^{\rm skew}(x,z) \\[3.mm]
\ds \delta_yK_{II}(x,z) & = & \ds \frac{K_{I}^0}{2}\left[ -2\varphi^0 + \frac{\partial \widetilde{\phi_y}}{\partial x}(x,z) \right]
- \frac{2-3\nu}{2-\nu}\frac{K_I^0}{2\pi} \, PV \int_{-\infty}^{+\infty}\frac{\widetilde{\phi_y}(x,z')-\widetilde{\phi_y}(x,z)}{(z'-z)^2}\,dz' \\[3.mm]
\ds \delta_yK_{III}(x,z) & = & \ds \frac{2(1-\nu)^2}{2-\nu}K_{I}^0\,\frac{\partial \widetilde{\phi_y}}{\partial z}(x,z)\ ; \\
\end{array}
\right.
\end{equation}
\begin{equation}\label{eqn:dyK1skew1}
\delta_y K_I^{\rm skew}(x,z) = \frac{\sqrt{2}}{4\pi} \, \frac{1-2\nu}{1-\nu} \, {\rm Re} \left\{\int_{-\infty}^{x} dx' \int_{-\infty}^{+\infty}
\frac{[K_{III}^0-i(1-\nu)K_{II}^0](\partial \widetilde{\phi_y}/\partial z)(x',z')}{(x-x')^{1/2}\left[x-x'+i(z-z')\right]^{3/2}} \, dz' \right\}.
\end{equation}
Accordingly, the expressions of the perturbed SIFs become for $x\geq 0$, by equations (\ref{eqn:AddPertSIF}), (\ref{eqn:dxK}) and (\ref{eqn:dyK1}):
\begin{equation}\label{eqn:Kpert}
\left\{
\begin{array}{lll}
K_{I}(x,z) & = & \ds K_I^0 + \frac{K_I^0}{2\pi} \, PV \int_{-\infty}^{+\infty}\frac{\phi_x(x,z')-\phi_x(x,z)}{(z'-z)^2}\,dz' \\[3.mm]
{} & {} & \ds - \frac{3}{2}K_{II}^0\frac{\partial \widetilde{\phi_y}}{\partial x}(x,z) - 2K_{III}^0\,\frac{\partial \widetilde{\phi_y}}{\partial z}(x,z) \\[3.mm]
{} & {} & \ds - \frac{K_{II}^0}{2\pi} \, PV \int_{-\infty}^{+\infty}\frac{\widetilde{\phi_y}(x,z')-\widetilde{\phi_y}(x,z)}{(z'-z)^2}\,dz'
+ \delta_y K_I^{\rm skew}(x,z) \\[3.mm]
K_{II}(x,z) & = & \ds - \frac{2}{2-\nu} K_{III}^0\,\frac{\partial \phi_x}{\partial z}(x,z)
+ \frac{2-3\nu}{2-\nu}\,\frac{K_{II}^0}{2\pi} \, PV \int_{-\infty}^{+\infty}\frac{\phi_x(x,z')-\phi_x(x,z)}{(z'-z)^2}\,dz' \\[3.mm]
{} & {} & \ds + \frac{K_{I}^0}{2}\,\frac{\partial \widetilde{\phi_y}}{\partial x}(x,z)
- \frac{2-3\nu}{2-\nu}\frac{K_I^0}{2\pi} \, PV \int_{-\infty}^{+\infty}\frac{\widetilde{\phi_y}(x,z')-\widetilde{\phi_y}(x,z)}{(z'-z)^2}\,dz' \\[3.mm]
K_{III}(x,z) & = & \ds K_{III}^0 + \frac{2(1-\nu)}{2-\nu} K_{II}^0\frac{\partial \phi_x}{\partial z}(x,z) \\[3.mm]
{} & {} & \ds + \frac{2+\nu}{2-\nu}\,\frac{K_{III}^0}{2\pi} \, PV \int_{-\infty}^{+\infty}\frac{\phi_x(x,z')-\phi_x(x,z)}{(z'-z)^2}\,dz' \\[3.mm]
{} & {} & \ds + \frac{2(1-\nu)^2}{2-\nu}K_{I}^0\,\frac{\partial \widetilde{\phi_y}}{\partial z}(x,z)
\end{array}
\right.
\end{equation}
where $\delta_y K_I^{\rm skew}(x,z)$ is given by equation (\ref{eqn:dyK1skew1}). Note that as a result of the general kink of the crack, in the expression of $K_{II}(x,z)$ here, the unperturbed SIF $K_{II}^0$ has cancelled with the term $-K_{I}^0\varphi^0 = - K_{II}^0$
in the expression of $\delta_yK_{II}(x,z)$ in (\ref{eqn:dyK1});
thus $K_{II}(x,z)$, unlike $K_{I}(x,z)$ and $K_{III}(x,z)$, is linear in the pair $(\phi_x,\widetilde{\phi_y})$ and vanishes when the perturbations $\phi_x$, $\widetilde{\phi_y}$ are zero. (This is of course because the conditions $\phi_x \equiv \widetilde{\phi_y} \equiv 0$ define the fundamental configuration of the crack satisfying the PLS).


\subsection{Definition of instability modes}\label{subsec:DefInstabModes}

The analysis will be based on consideration of instability modes consisting of perturbations of the crack front and surface of the following form:
\begin{itemize}
\item In the region $x\geq 0$:
\begin{equation}\label{eqn:ComplexInstMode}
\left\{
\begin{array}{lll}
\phi_x(x,z) & = & \mbox{Re}\left[ e^{\lambda x} \psi_x(z) \right] \\
\widetilde{\phi_y}(x,z) & = & \mbox{Re}\left[ e^{\lambda x} \psi_y(z) \right]
\end{array}
\right.
\end{equation}
where $\lambda$ is an unknown complex scalar, the ``complex growth rate'' of the instability mode, and $\psi_x(z)$, $\psi_y(z)$ unknown complex functions. This is the same form as that envisaged in our previous works on crack propagation in mixed-mode I+III \citep{LKL11,LKPV18}, except that $\lambda$ was assumed to be real there; such an assumption will be seen to no longer be acceptable in the presence of global mode II ($K_{II}^0\neq 0$).
\item In the region $x<0$:

\noindent No assumption is made on the values $\phi_x$ and $\widetilde{\phi_y}$ other than their boundedness. (It will be shown in the sequel that they have no impact whatsoever upon the stability analyzes).\footnote{The different assumptions made in the regions $x\geq 0$ and $x<0$ are consequences of the fact that the exponential variation of the crack perturbation basically arises from the double propagation criterion assumed to be obeyed during the mixed-mode propagation of the crack. This criterion can be applied in the region $x\geq 0$ resulting from such a propagation, but not in the region $x<0$ corresponding to the initial flat crack generated in some other way.}
\end{itemize}

Repeated use will be made in the sequel of the following property, tied to the non-vanishing of the imaginary part of the number $\lambda$:

$({\mathcal P})$: {\it If $A$ is a complex number such that ${\rm Re}(Ae^{\lambda x})=0$ for every non-negative real number $x$, then necessarily $A=0$}.

The natural invariance of the problem in the direction $z$ of the crack front suggests using Fourier transforms in this direction. The definition adopted here for the Fourier transform $\widehat{\chi}(k)$ of an arbitrary function $\chi(z)$ is
\begin{equation}\label{eqn:DefFourier}
\chi(z) = \int_{-\infty}^{+\infty} \widehat{\chi}(k) \,e^{ikz} dk \quad \Leftrightarrow \quad
\widehat{\chi}(k) = \frac{1}{2\pi} \int_{-\infty}^{+\infty} \chi(z) \,e^{-ikz} dz.
\end{equation}
We then define a dimensionless ``normalized complex growth rate'' $\xi$ of each Fourier component of the instability mode by the formula
\begin{equation}\label{eqn:DefXi}
\xi \equiv \frac{\lambda}{|k|}
\end{equation}
which ``compares'' its complex growth rate $\lambda$ in the direction $x$ to its wavenumber $|k|$ in the direction $z$.


\section{Linear stability analysis for small values of $K_{III}$ and variable $G_\mathrm{c}$}\label{sec:StabAnal1}

\subsection{Additional hypothesis and resulting simplifications}\label{subsec:HypAnal1}

In a first step, we wish to present a fully rigorous linear stability analysis devoid of any approximation. The following problem then arises. A solution of the bifurcation problem varying exponentially with $x$, as looked for here, is possible only provided all terms in the expressions of the perturbations of the SIFs vary exponentially themselves. But this obviously cannot be true of the term $\delta_y K_I^{\rm skew}(x,z)$ given by equation (\ref{eqn:dyK1skew1}), in the absence of any specific assumption on the variation of $\widetilde{\phi_y}$ in the region $x<0$ prior to propagation of the crack in mode I+II+III.

To obviate this difficulty, we introduce the additional assumption that {\it the ratio $\rho^0=K_{III}^0/K_{I}^0$ is much smaller than unity} like $|\varphi^0|=|K_{II}^0|/K_{I}^0$; accordingly, the treatment will be limited to first order in the pair $(\varphi^0,\rho^0)$. This hypothesis is not overly restrictive since a large number of fracture experiments in mixed-mode I+III or I+II+III have been performed under such conditions. Note that it does not enforce any restriction on the ratio $R^0=\varphi^0/\rho^0=K_{II}^0/K_{III}^0$ which may still take arbitrary values.

The advantage of this extra hypothesis is as follows. Using the PLS with the expression (\ref{eqn:Kpert})$_2$ of $K_{II}(x,z)$, one easily sees that $\widetilde{\phi_y}$ is of the order of $\phi_x$ times a term of first order in the pair $(\varphi^0,\rho^0)$. Examination of the expression (\ref{eqn:dyK1skew1}) of $\delta_y K_I^{\rm skew}(x,z)$ then reveals that the integrand there is of second order in this pair;\footnote{The PLS applies only in the region $x\geq 0$, but the observation that the integrand of $\delta_y K_I^{\rm skew}(x,z)$ is of second order in $(\varphi^0,\rho^0)$ holds even at points $(x'<0,z')$, since it is of this order at $x'=0$ (as a consequence of the PLS) and the perturbation $\widetilde{\phi_y}$ is assumed to be bounded in the region $x'<0$.} it follows that $\delta_y K_I^{\rm skew}(x,z)$ is itself of second order and therefore negligible, which settles the issue raised by this term.

An ancillary consequence of the additional hypothesis of smallness of $\rho^0$ is that in the expression (\ref{eqn:Kpert})$_1$ of the perturbed SIF $K_{I}(x,z)$, the other terms involving $\widetilde{\phi_y}$ are also negligible. This expression therefore simply reduces, for every non-negative $x$, to
\begin{equation}\label{eqn:K1pert}
K_{I}(x,z) = K_I^0 + \frac{K_I^0}{2\pi} \, PV \int_{-\infty}^{+\infty}\frac{\phi_x(x,z')-\phi_x(x,z)}{(z'-z)^2}\,dz'.
\end{equation}
While no extra simplifications are possible in the expressions (\ref{eqn:Kpert})$_{2,3}$ of the perturbed SIFs $K_{II}(x,z)$, $K_{III}(x,z)$, those perturbed SIFs, like $K_{I}(x,z)$ in (\ref{eqn:K1pert}), depend only on the instantaneous crack front shape, thereby making the shape of the fracture surface for $x<0$ irrelevant for the linear stability problem in the stated limit $\rho^0\ll 1$.


\subsection{Application of the principle of local symmetry}\label{subsec:ApplicPLSAnal1}

Using the definition (\ref{eqn:ComplexInstMode}) of the instability modes and expressing $\psi_x$, $\psi_y$ in terms of their Fourier transforms $\widehat{\psi_x}$, $\widehat{\psi_y}$, one gets from formula (\ref{eqn:Kpert})$_2$, for every non-negative $x$,
\begin{equation}\label{eqn:K2Pert}
\begin{array}{lll}
\ds \frac{K_{II}(x,z)}{K_I^0} & = & \ds {\rm Re}\left\{ e^{\lambda x} \int_{-\infty}^{+\infty} \left[ - \left( \frac{2i}{2-\nu}\,\rho^0k + \frac{2-3\nu}{2(2-\nu)}\,\varphi^0|k| \right)\widehat{\psi_x}(k) \right. \right. \\[3mm]
{} & {} & \ds \left. \left. + \left( \frac{\lambda}{2} + \frac{2-3\nu}{2(2-\nu)}|k| \right)\widehat{\psi_y}(k) \right] e^{ikz}dk \right\}.
\end{array}
\end{equation}
This expression must be equated to zero for every $x\geq 0$ according to the PLS. It then follows from property $({\mathcal P})$ of Subsection \ref{subsec:DefInstabModes} that the Fourier transforms $\widehat{\psi_x}$ and $\widehat{\psi_x}$ are necessarily connected through the relation
\begin{equation}\label{eqn:PhiyVsPhix}
\widehat{\psi_y}(k) = \frac{(2-3\nu)\varphi^0+4i\rho^0\,{\rm sgn}(k)}{2-3\nu+(2-\nu)\xi}\, \widehat{\psi_x}(k)
\end{equation}
where ${\rm sgn}(k)$ denotes the sign of $k$ and $\xi$ is defined by equation (\ref{eqn:DefXi}).


\subsection{Application of \cite{G20}'s criterion}\label{subsec:ApplicGriffithAnal1}

To apply \cite{G20}'s criterion with a local $G_\mathrm{c}$ given by equation (\ref{eqn:GcVsRho}), the first task is to express the perturbations $\delta K_I(x,z)$, $\delta K_{III}(x,z)$ in terms of the Fourier transforms $\widehat{\psi_x}(k)$, $\widehat{\psi_y}(k)$. One gets from equations (\ref{eqn:K1pert}) and (\ref{eqn:Kpert})$_3$, for every non-negative $x$,
\begin{equation}\label{eqn:PertK13}
\left\{
\begin{array}{lll}
\ds \frac{\delta K_I(x,z)}{K_I^0} & = & \ds - {\rm Re} \left\{ e^{\lambda x} \int_{-\infty}^{+\infty}
\frac{|k|}{2}\,\widehat{\psi_x}(k) e^{ikz}dk \right\} \\[5mm]
\ds \frac{\delta K_ {III}(x,z)}{K_I^0} & = & \ds {\rm Re} \left\{ \frac{e^{\lambda x}}{2-\nu} \times \right. \\[3mm]
{} & {} & \ds \left. \int_{-\infty}^{+\infty} \left[
\left( 2i(1-\nu)\varphi^0k - \frac{2+\nu}{2}\,\rho^0|k| \right)
\widehat{\psi_x}(k) + 2i\left(1-\nu\right)^2 k \widehat{\psi_y}(k) \right] e^{ikz}dk \right\}.
\end{array}
\right.
\end{equation}

One may then calculate the perturbations $\delta G$ of the energy-release-rate $G$, and $\delta\rho$ of the ratio $\rho=K_{III}/K_I$:
\begin{itemize}
\item With the approximations made, the unperturbed value of the energy-release-rate is $G^0 = \frac{1-\nu^2}{E}(K_I^0)^2$ and its perturbation is $\delta G(x,z) = 2\,\frac{1-\nu^2}{E}\, K_I^0 \delta K_I(x,z)$. Using equation (\ref{eqn:PertK13})$_1$, one then gets for $x\geq 0$,
\begin{equation}\label{eqn:DeltaG}
\frac{\delta G(x,z)}{G^0} = 2\, \frac{\delta K_I(x,z)}{K_I^0} = - {\rm Re}\left\{ e^{\lambda x} \int_{-\infty}^{+\infty}|k|\widehat{\psi_x}(k) e^{ikz}dk \right\}.
\end{equation}
\item One similarly gets for $x\geq 0$, after a lengthy calculation based on equations (\ref{eqn:PertK13}) and the relation (\ref{eqn:PhiyVsPhix}) connecting $\widehat{\psi_x}$ and $\widehat{\psi_y}$:
\begin{equation}\label{eqn:DeltaRho}
\begin{array}{lll}
\delta \rho(x,z) & = & \ds {\rm Re}\left\{ e^{\lambda x} \int_{-\infty}^{+\infty} \frac{-(4-5\nu+\nu\lambda/|k|)\rho^0 + 2(1-\nu)(2-3\nu+\lambda/|k|)i\varphi^0{\rm sgn}(k)}{2-3\nu+(2-\nu)\lambda/|k|} \times \right. \\[3mm]
{} & {} & \ds |k|\widehat{\psi_x}(k) e^{ikz}dk \bigg \}.
\end{array}
\end{equation}
\end{itemize}

Then Griffith's criterion
\begin{equation*}
G(x,z) = G^0 + \delta G(x,z) = G_\mathrm{c}(\rho^0) + \frac{dG_\mathrm{c}}{d\rho}(\rho^0)\,\delta\rho(x,z),
\end{equation*}
applied in the region $x\geq 0$ corresponding to propagation of the crack in mixed-mode I+II+III, yields at the various orders in the pair $({\phi_x},\widetilde{\phi_y})$:
\begin{itemize}
\item At order 0:
\begin{equation*}
G^0 = G_\mathrm{c}(\rho^0) \quad (=G_\mathrm{cI}[1 + \gamma (\rho^0)^{\kappa}]).
\end{equation*}
\item At order 1: for every $x\geq 0$:
\begin{equation*}
\frac{\delta G(x,z)}{G^0} = \frac{d({\rm ln}G_\mathrm{c})}{d\rho}(\rho^0)\,\delta\rho(x,z).
\end{equation*}
One then concludes from equations (\ref{eqn:DeltaG}) and (\ref{eqn:DeltaRho}), using the expression (\ref{eqn:GcVsRho}) of $G_\mathrm{c}$ and property $({\mathcal P})$ of Subsection \ref{subsec:DefInstabModes}, that if $\widehat{\psi_x}(k) \neq 0$ (non-trivial instability mode), then necessarily
\begin{equation}\label{eqn:xiAnal1}
\xi = \frac{N_1+iN_2}{D_1+iD_2}
\ , \
\left\{
\begin{array}{lll}
N_1 & \equiv & -2+3\nu+(4-5\nu)X \\
N_2 & \equiv & \ds -2(1-\nu)(2-3\nu)XR^0s \\
D_1 & \equiv & 2-\nu-\nu X \\
D_2 & \equiv & \ds 2(1-\nu)XR^0s
\end{array}
\right.
\ , \
\left\{
\begin{array}{lll}
X & \equiv & \ds \frac{\kappa\gamma(\rho^0)^{\kappa}}{1+\gamma(\rho^0)^{\kappa}} \\
s & \equiv & {\rm sgn}(k).
\end{array}
\right.
\end{equation}
\end{itemize}
The quantities $N_1$, $N_2$, $D_1$, $D_2$ and $X$ here depend upon the parameters $\varphi^0$ and $\rho^0$ (or $R^0=\varphi^0/\rho^0$), although this is omitted in the notation to keep it reasonably light.


\subsection{Condition for incipient instability}\label{subsec:IncipInstabAnal1}

In contrast to what occurred in the case of mixed-mode I+III without mode II envisaged by \cite{LKL11,LKPV18}, the normalized growth rate $\xi=\lambda/|k|$ of the perturbation is now no longer real but {\it complex}:
\begin{equation}\label{eqn:RealImPartsXi}
\xi \equiv \xi_1 + i\xi_2 \quad , \quad
\left\{
\begin{array}{lll}
\xi_1 & \equiv & \ds \frac{N_1D_1+N_2D_2}{D_1^2+D_2^2} \\[3mm]
\xi_2 & \equiv & \ds \frac{N_2D_1-N_1D_2}{D_1^2+D_2^2}\,.
\end{array}
\right.
\end{equation}
The condition for incipient instability of coplanar propagation then reads:
\begin{equation}\label{eqn:CondIncipInstab}
\begin{array}{c}
{\rm Re}\,\xi \equiv \xi_1 = 0 \quad \Leftrightarrow \\[3mm]
\ds N_1D_1+N_2D_2 = \left[ -2+3\nu+(4-5\nu)X \right](2-\nu-\nu X) - 4(1-\nu)^2(2-3\nu)(R^0)^2X^2 = 0
\end{array}
\end{equation}
where the expressions (\ref{eqn:xiAnal1})$_{2-5}$ of $N_1$, $N_2$, $D_1$, $D_2$ have been used. For any given value of the ratio $R^0 = {K_{II}^0}/{K_{III}^0}$, equation (\ref{eqn:CondIncipInstab}), with $X$ given by equation (\ref{eqn:xiAnal1})$_6$, is equivalent to an algebraic equation of the second degree on the unknown $(\rho^0)^{\kappa}$, which determines the critical value $\rho^{\rm cr}$ of the ratio $\rho^0=K_{III}^0/K_{I}^0$ leading to incipient instability. (Note that this value must be small to ensure self-consistency of the treatment based on the hypotheses $|\varphi^0|\ll 1$, $\rho^0\ll 1$).

We now analyze the predictions of equation (\ref{eqn:CondIncipInstab}) in more detail. The first remark is that for $\rho^0 = 0$, $X = 0$ so that $N_1D_1+N_2D_2$ takes the negative value $(-2+3\nu)(2-\nu)$. Thus by equation (\ref{eqn:RealImPartsXi})$_2$, $\xi_1$ is negative - implying configurational stability of the propagating crack - for small values of $\rho^0$. Now by definition, the critical value $\rho^{\rm cr}$ is the {\it smallest} positive solution in $\rho^0$ of equation (\ref{eqn:CondIncipInstab}), ensuring that $\xi_1=0$. It follows that {\it $\xi_1$ is negative - implying stability - as long as $\rho^0$ remains smaller than the critical value $\rho^{\rm cr}$}, vanishes when $\rho^0$ reaches it, and {\it becomes positive - implying instability - when $\rho^0$ becomes larger}.

Another remark pertains to the qualitative influence of the ratio $R^0$ upon the threshold $\rho^{\rm cr}$. We have just seen that $N_1D_1+N_2D_2$ is negative for $0\leq \rho^0<\rho^{\rm cr}$. Now for a given $\rho^0$, $N_1D_1+N_2D_2$ decreases when $|R^0|$ increases; hence when this absolute value increases, the interval $[0,\rho^{\rm cr})$ over which $N_1D_1+N_2D_2$ is negative can only become larger, implying that $\rho^{\rm cr}$ can only increase. In other words, {\it presence of mode II necessarily leads to an increase of the critical value $\rho^{\rm cr}$ of the ratio ${K_{III}^0}/{K_{I}^0}$ corresponding to incipient instability}.


\subsection{Geometry of instability modes}\label{subsec:GeomModeAnal1}

To discuss instability modes, we consider a positive\footnote{The mode obtained for the negative value $-k$ is readily checked to be identical.} value of the wavenumber, $k$, and a Fourier transform $\widehat{\psi_x}(k')$ of the form
\begin{equation*}
\widehat{\psi_x}(k') = A_x\,e^{i\theta}\delta(k'-k)
\end{equation*}
where $A_x$ and $\theta$ are real numbers and $\delta$ Dirac's generalized function. This Fourier transform corresponds to a function $\psi_x(z)$ of the form
\begin{equation*}
\psi_x(z) = A_x\,e^{i(kz+\theta)}.
\end{equation*}
The normalized complex growth rate $\xi$ is fixed, independently of the value of $k$, by equations (\ref{eqn:RealImPartsXi}). The complex growth rate $\lambda$ is then given by $\lambda = \xi |k| = (\xi_1+i\xi_2)k$ and the in-plane perturbation $\phi_x(x,z)$ by
\begin{equation}\label{eqn:InstabModeX}
\phi_x(x,z) = A_x\, e^{\xi_1kx}\cos\left[ k(z+\xi_2x)+\theta \right] ,
\end{equation}
see equation (\ref{eqn:ComplexInstMode})$_1$.

To calculate the out-of-plane perturbation $\widetilde{\phi_y}$, split the ratio connecting the Fourier transforms $\widehat{\psi_x}$, $\widehat{\psi_y}$ in equation (\ref{eqn:PhiyVsPhix}) into real and imaginary parts:
\begin{equation}\label{eqn:DefPQ}
\frac{(2-3\nu)\varphi^0+4i\rho^0}{2-3\nu+(2-\nu)(\xi_1+i\xi_2)} \equiv p+iq.
\end{equation}
The precise expressions of the quantities $p$ and $q$ here do not matter; it suffices that they are of first order in the pair $(\varphi^0,\rho^0)$ and therefore small. With this notation,
\begin{equation*}
\widehat{\psi_y}(k') = (p+iq)A_x\,e^{i\theta}\delta(k'-k) \quad \Rightarrow \quad \psi_y(z) = (p+iq)A_x\,e^{i(kz+\theta)}.
\end{equation*}
It then follows from equation (\ref{eqn:ComplexInstMode})$_2$ that
\begin{equation}\label{eqn:InstabModeY}
\begin{array}{lll}
\widetilde{\phi_y}(x,z) & = & \mbox{Re}\left[ e^{(\xi_1+i\xi_2)kx} (p+iq)A_x\,e^{i(kz+\theta)} \right] \\
{} & = & A_x\, e^{\xi_1kx} \left\{ p\,\cos\left[ k(z+\xi_2x)+\theta \right] - q\,\sin\left[ k(z+\xi_2x)+\theta \right] \right\}.
\end{array}
\end{equation}

\begin{figure}[h]
\centerline{\includegraphics[height=5cm]{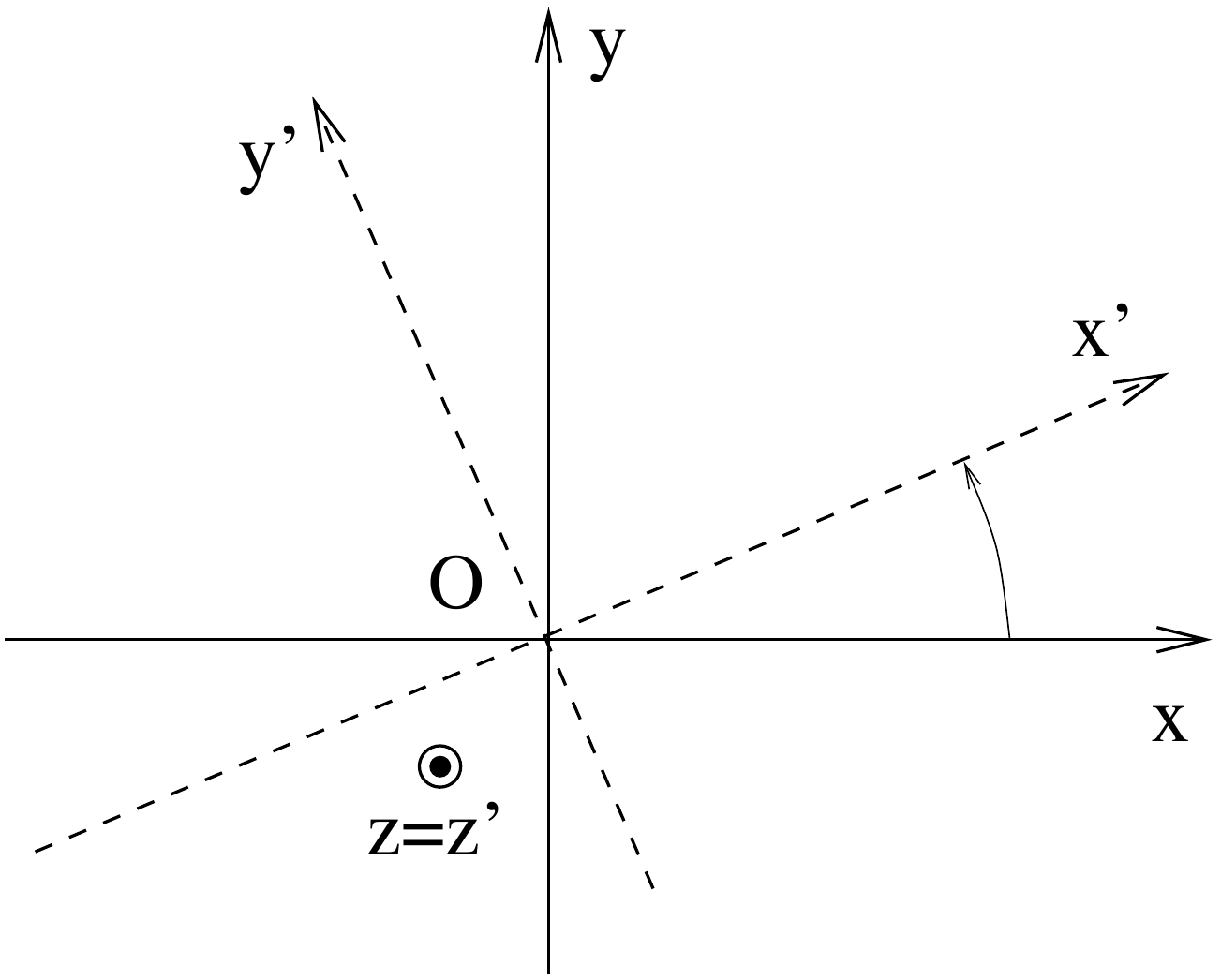}} \caption{Change of frame in the plane $Oxy$ orthogonal to the unperturbed crack front.}
\label{fig:ChgmtFrame}
\end{figure}

The geometrical interpretation of the instability mode defined by equations (\ref{eqn:InstabModeX}) and (\ref{eqn:InstabModeY}) is made easier by introducing a new frame $(Ox'y'z)$ obtained through rotation of the original one $(Oxyz)$ by a small angle $\eta$ about the axis $Oz$ (Fig. \ref{fig:ChgmtFrame}).

The perturbation vector of the front, $\phi_x{\bf e}_x+\widetilde{\phi_y}{\bf e}_y$, may be expressed in the new frame as ${\phi_x}'{{\bf e}_x}'+\widetilde{\phi_y}'{{\bf e}_y}'$ where, to first order in $\eta$,
\begin{equation*}
\left\{
\begin{array}{lll}
{\phi_x}'(x,z) & = & {\phi_x}(x,z) + \eta \widetilde{\phi_y}(x,z) \\
\widetilde{\phi_y}'(x,z) & = & - \eta{\phi_x}(x,z) + \widetilde{\phi_y}(x,z).
\end{array}
\right.
\end{equation*}
The cosine term in $\widetilde{\phi_y}'$ may be eliminated by ascribing $\eta$ the (small) value $p$; the expressions of ${\phi_x}'$ and $\widetilde{\phi_y}'$ then become, to first order in the pair $(\varphi^0,\rho^0)$:
\begin{equation}\label{eqn:InstabModeXY}
\left\{
\begin{array}{lll}
{\phi_x}'(x,z) & = & A_x\, e^{\xi_1kx}\cos\left[ k(z+\xi_2x)+\theta \right] \\
\widetilde{\phi_y}'(x,z) & = & A_y\, e^{\xi_1kx}\sin\left[ k(z+\xi_2x)+\theta \right]
\end{array}
\right.
\quad {\rm where} \quad \frac{A_y}{A_x} \equiv - q.
\end{equation}

Like in the case of a mixed-mode I+III loading envisaged by \citep{LKL11,LKPV18}, equation (\ref{eqn:InstabModeXY}) defines a perturbed crack front having the shape of an elliptic helix, of central axis coinciding with the unperturbed front, and semi-axes growing in proportion and exponentially with the distance $x$ of propagation. There are however two novelties:
\begin{itemize}
\item The presence of mode II induces a small rotation of the principal axes of the ellipse (projection of the helix onto the plane $Oxy$) about the direction $z$ of the unperturbed crack front.
\item More importantly, the helix no longer moves in the general direction $x$ of propagation of the crack, but {\it drifts along the front as it propagates}, with a ``drift velocity'' given by
\begin{equation}\label{eqn:LaterVeloc}
\frac{dz}{dx} = - \xi_2 = \frac{N_1D_2-N_2D_1}{D_1^2+D_2^2}
= \frac{ 2(1-\nu)^2[ 2-3\nu+(4-3\nu)X]XR^0 }{ (2-\nu-\nu X)^2 + 4(1-\nu)^2X^2(R^0)^2 }
\end{equation}
where the expression of $N_1D_2-N_2D_1$ has been developed and rearranged. Note that this drift velocity depends on both ratios $R^0 = K_{II}^0/K_{III}^0$ {\it and} $\rho^0=K_{III}^0/K_{I}^0$ (through the parameter $X$ defined in Eq.~\eqref{eqn:xiAnal1}).
\end{itemize}

The drifting motion of the instability modes is illustrated schematically in Fig. \ref{fig:CrabMotion}(b), with a comparison in Fig. \ref{fig:CrabMotion}(a) with the case of a mixed-mode I+III loading for which such a motion is absent. Note that when both $K_{II}^0$ and $K_{III}^0$ are positive, the ``drift angle'' $\alpha = \arctan(dz/dx)$ is positive too, as illustrated in Fig. \ref{fig:CrabMotion}(b). In contrast, if $K_{II}^0$ and $K_{III}^0$ are of opposite sign, the drift angle is negative.


\begin{figure}[h]
\centering
{
\includegraphics[width=0.45\textwidth, trim = 25mm 30mm 25mm 30 mm, clip = true]{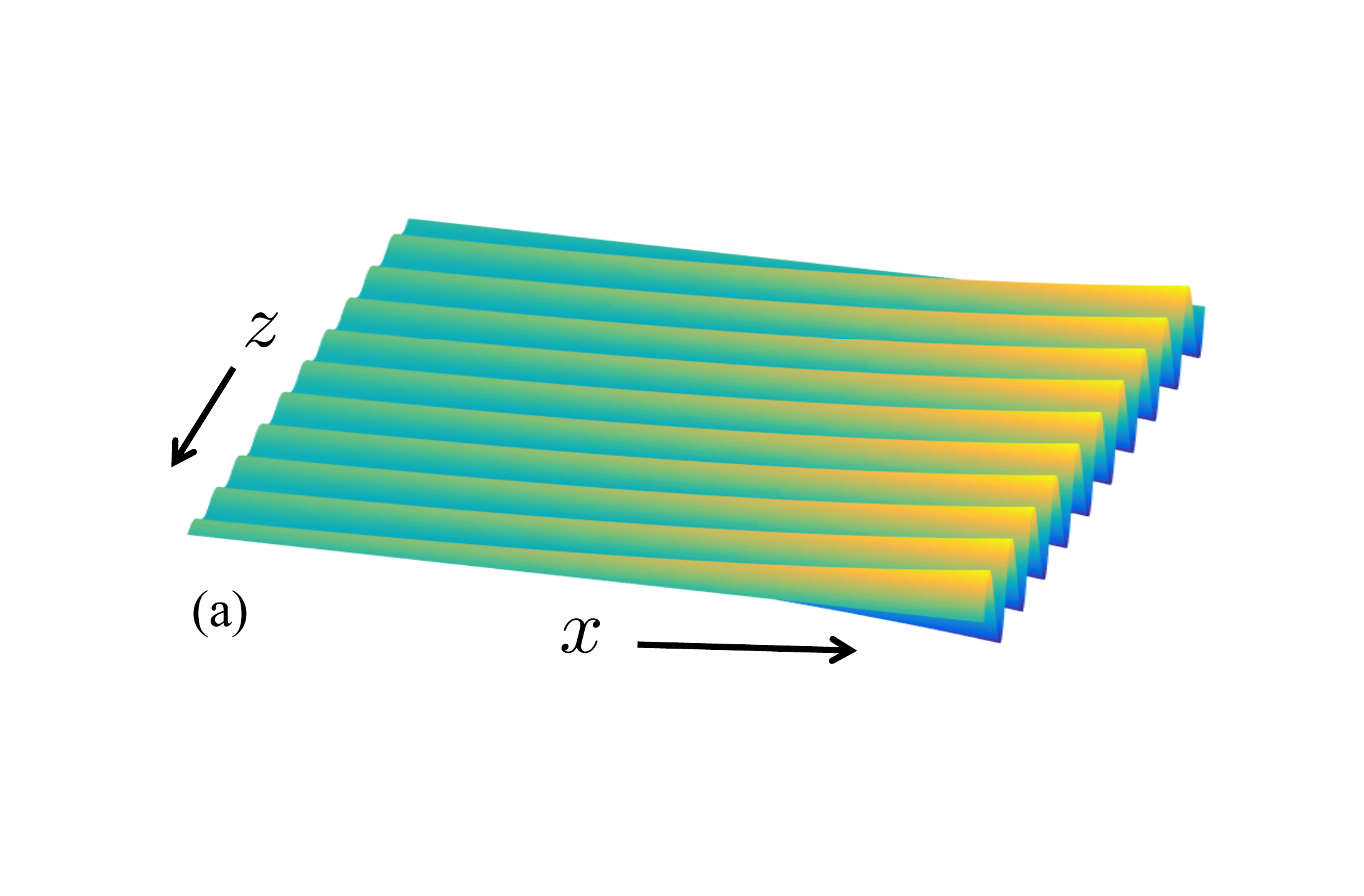}
}
{
\includegraphics[width=0.45\textwidth, trim = 25mm 30mm 25mm 30 mm, clip = true]{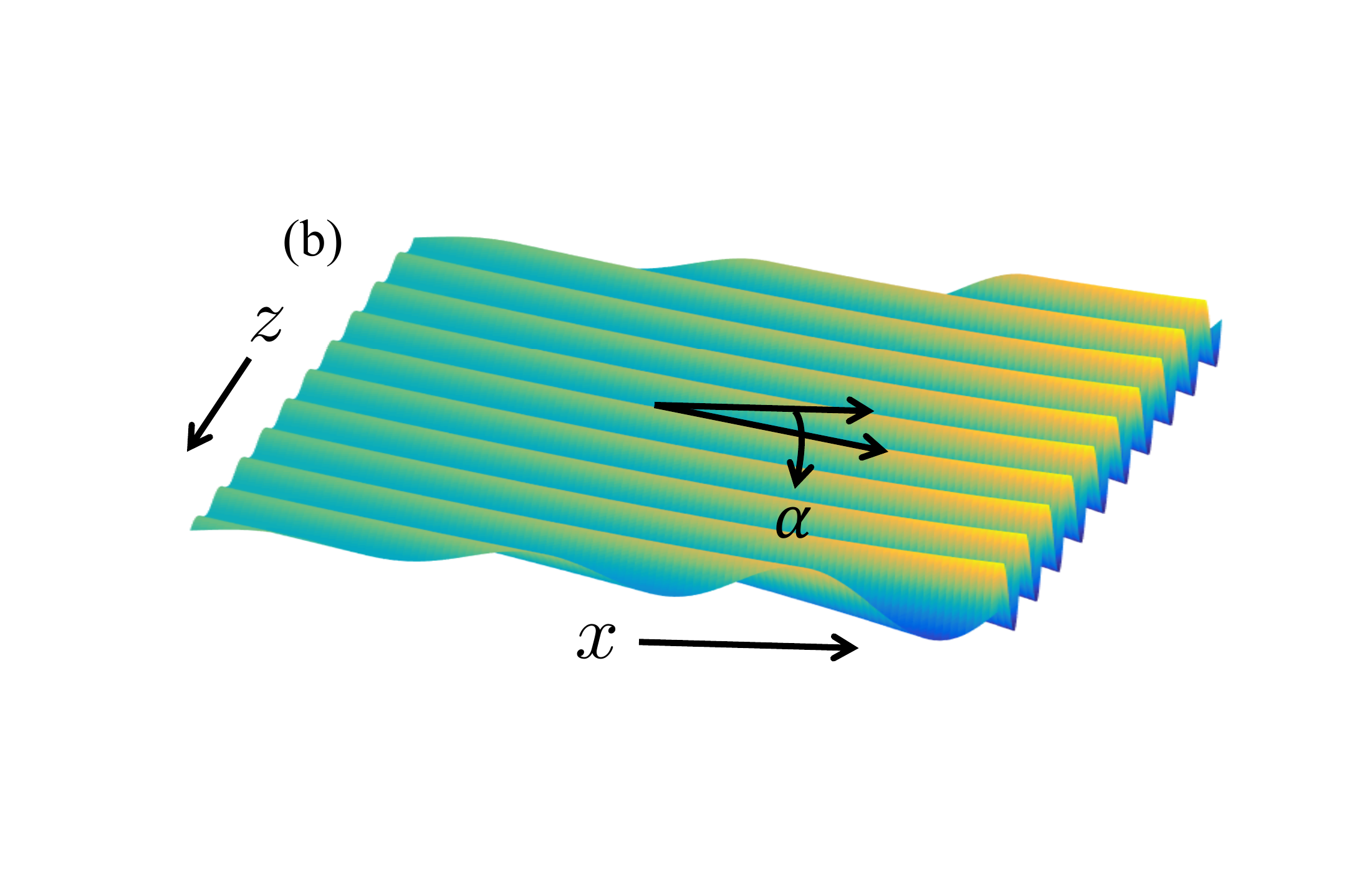}
}
\caption{\label{fig:CrabMotion} Geometry of the instability modes (a) under mixed-mode I + III and (b) under mixed-mode I + II + III. In the presence of a mode II component, the facets drift along the crack front with an angle $\alpha$ with respect to the propagation direction $x$, that are also remarkably similar to experimental observations (see Fig. \ref{fig:ExpSetup} (d)). The sign of $\alpha$ is set by that of the ratio $K_{II}^0/K_{III}^0$ and is positive when both shear loading components are of identical sign.}
\label{fig:CrabMotion}
\end{figure}

It is interesting to note that in the present analysis, the drifting motion of facets along the crack front results from combination of existence of a mode II loading component ($K_{II}^0\neq 0$) {\it and} dependence of the toughness $G_\mathrm{c}$ upon the ratio $K_{III}/K_{I}$. Indeed the drift velocity, being proportional to both $R^0$ and $X$, is zero either if $K_{II}^0=0$, or if $G_\mathrm{c}$ is independent of $\rho$, as $\gamma=0$ implies $X=0$.

It may also be noted that the existence of a drift of facets formed by crack front fragmentation, albeit not its direction along $z$, could somehow be expected, as the additional mode II component breaks the invariance of the problem in a rotation of the geometry and loading of $180^\circ$ around the $x$-axis,\footnote{Such a rotation leaves the mode I and III loading components unchanged, but changes the sign of the mode II component.} and the ensuing symmetry $z \rightarrow -z$ along the crack front. However this symmetry argument is independent of the possible dependence of $G_\mathrm{c}$ upon $K_{III}/K_{I}$; hence it comes as somewhat of a surprise that the present analysis concludes that the fracture pattern produced by fragmentation actually breaks the symmetry $z \rightarrow -z$ {\it only in the presence of such a dependence}. This issue will be discussed in more detail in Section \ref{sec:StabAnal2} below.


\subsection{Numerical illustrations}\label{sec:NumExAnal1}

We shall now numerically illustrate the predictions of the preceding stability analysis by considering various values of the material parameters involved in the model. (The comparison between theoretical predictions and experimental observations will be envisaged in the discussion Section \ref{sec:Discuss} from a purely qualitative viewpoint, and from a quantitative viewpoint in some future work). The values of the ratio $\varphi^0=K_{II}^0/K_{I}^0$ considered will not exceed $10~\%$, in line with the hypothesis of small mode II introduced in Subsection \ref{subsec:HypGeomLoad}.

\begin{figure}[h]
\centerline{\includegraphics[height=6cm]{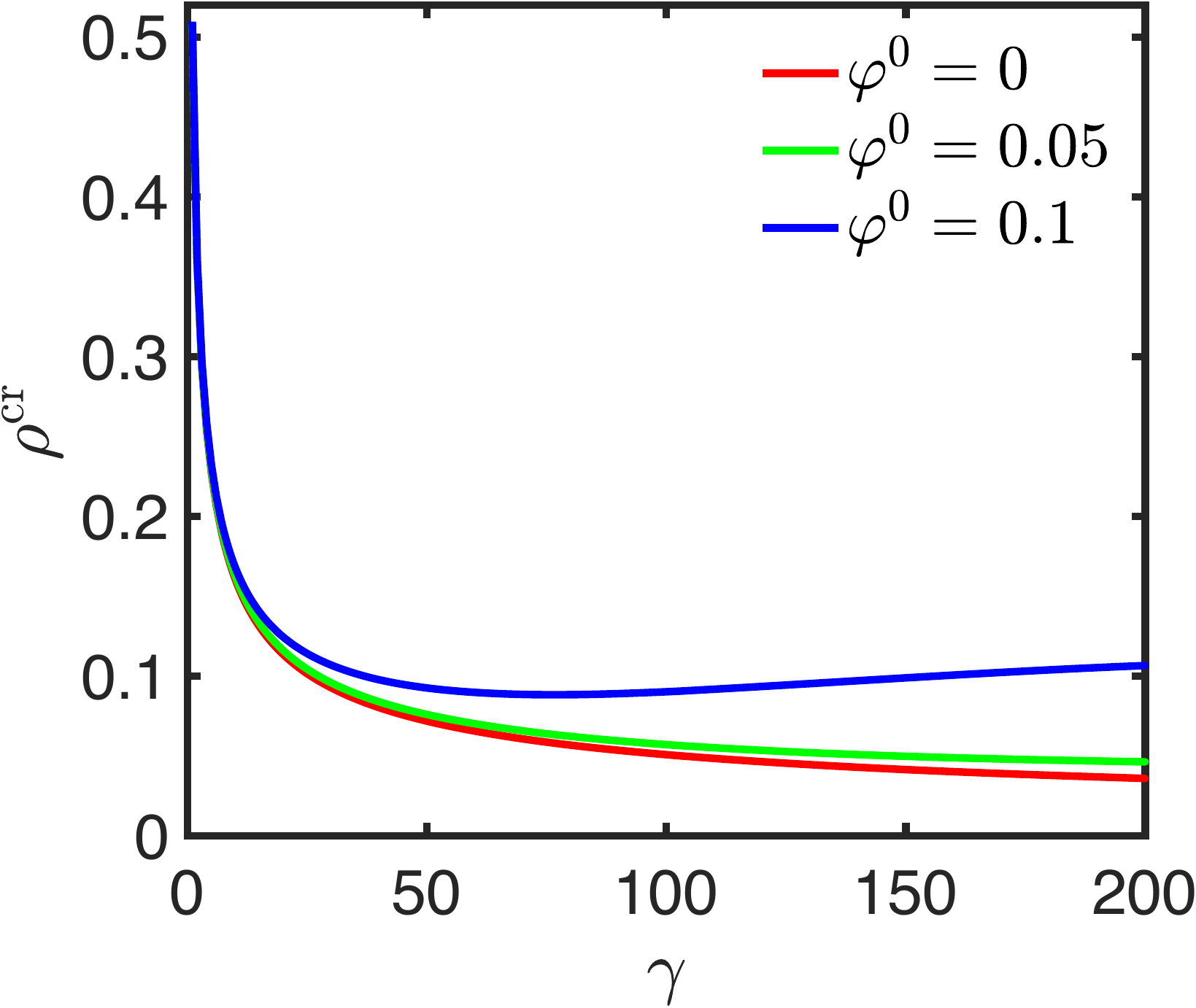}} \caption{Instability threshold $\rho^{\rm cr}$ as a function of the material parameter $\gamma$ specifying the mode III-induced toughening, for $\nu=0.38$, $\kappa=2$ and three values of $\varphi^0=K_{II}^0/K_{I}^0$.}
\label{fig:RhocVsGamma}
\end{figure}

Figure~\ref{fig:RhocVsGamma} first shows the instability threshold $\rho^{\rm cr}$ as a function of the ``toughening parameter'' $\gamma$ for various values of the ratio $\varphi^0=K_{II}^0/K_{I}^0$. Here, we choose the value $\nu=0.38$ for Poisson's ratio, which is typical for PMMA. We also ascribe the parameter $\kappa$ a value of $2$, the lowest one simultaneously ensuring parity and regularity of the function $G_\mathrm{c}(\rho)$. The increase of the threshold arising from the presence of mode II is conspicuous, all the more so for large values of $\gamma$.

\begin{figure}[h]
\centering
{
\includegraphics[height=5.5cm]{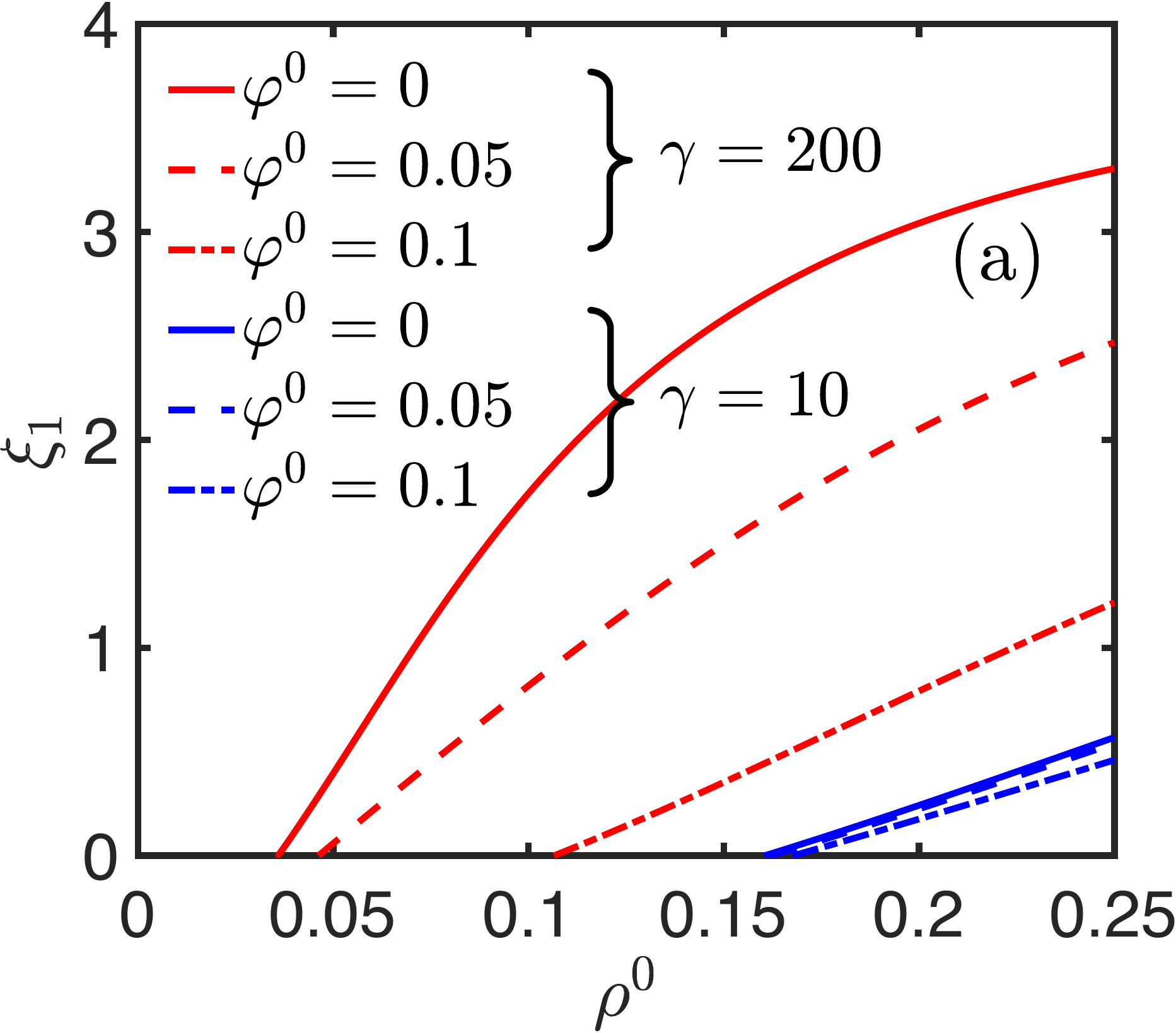}
}
{
\includegraphics[height=5.6cm]{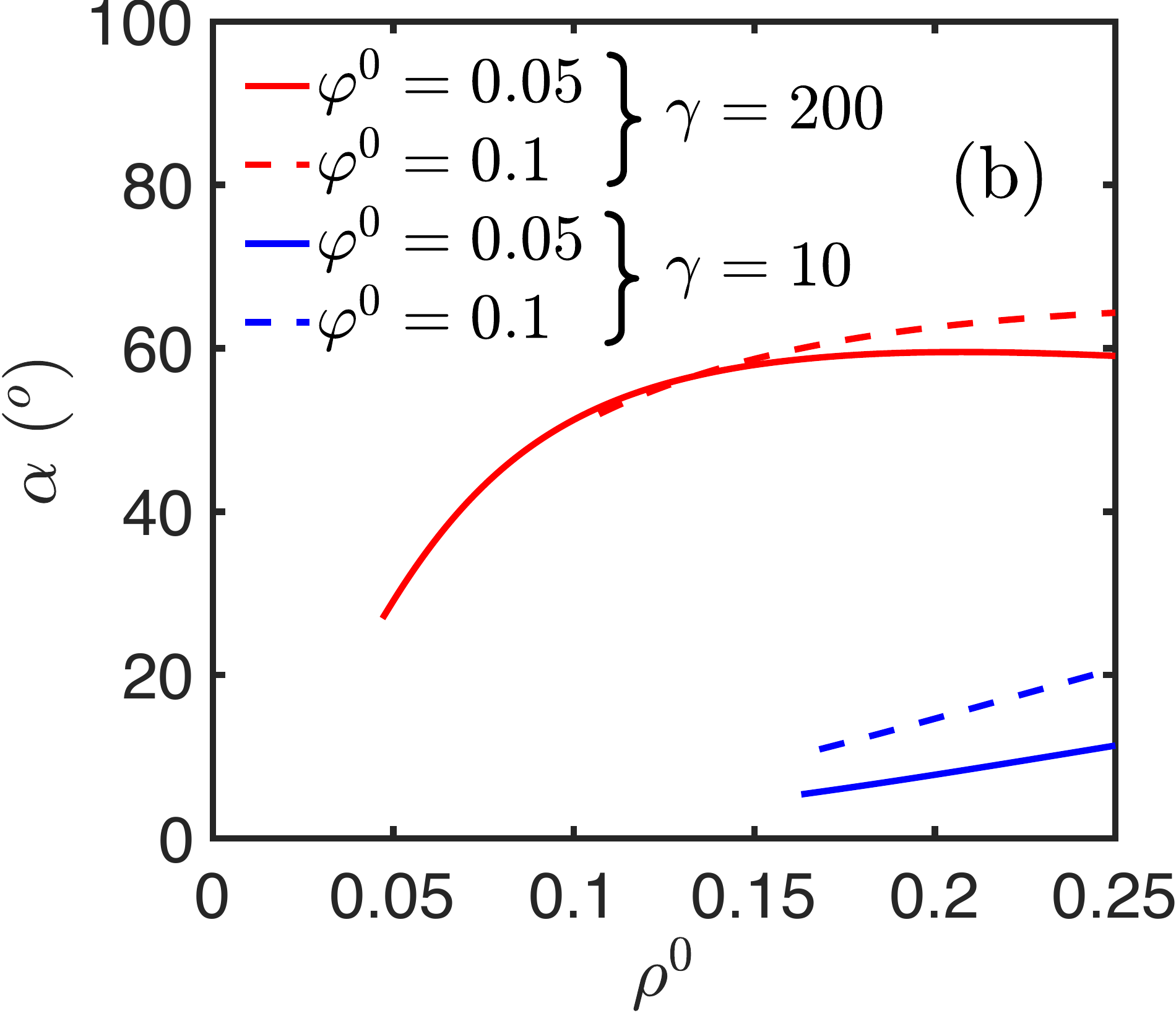}
}
\caption{\label{fig:ReImXi} Growth rate $\xi_1 = \mbox{Re}\ \xi$ and drift angle $\alpha = - \arctan(\mbox{Im}\ \xi)$ of the instability mode as functions\emph{} of the ratio $\rho^0=K_{III}^0/K_{I}^0$, for $\nu=0.38$, $\kappa=2$ and several values of $\varphi^0=K_{II}^0/K_{I}^0$.}
\label{fig:DzDx}
\end{figure}

Figure~\ref{fig:ReImXi} shows the real and imaginary parts $\xi_1$, $\xi_2$ of the normalized growth rate $\xi=\lambda/|k|$ of the instability modes, as functions of the ratio $\rho^0=K_{III}^0/K_{I}^0$, for $\nu=0.38$, $\kappa=2$ and several values of the ratio $\varphi^0=K_{II}^0/K_{I}^0$. Two values of the toughening parameter are considered: $\gamma = 10$, for a material with moderately mode III-dependent fracture energy, and $\gamma = 200$, for a material with highly mode III-dependent fracture energy. The results are displayed only for $\rho > \rho^{\rm cr}$ since there is no instability for $\rho < \rho^{\rm cr}$. Figure~\ref{fig:DzDx}(a) confirms that application of a mode II component stabilizes the crack front, since the parameter $\xi_1$ characterizing the exponential growth rate of the perturbation during propagation decreases when $\varphi^0$ increases. Figure~\ref{fig:DzDx}(b) shows that the drift velocity $dz/dx$ increases with both the amount of in-plane shear $K_{II}^0$ and the parameter $\gamma$. It also increases with the amount of anti-plane shear $K_{III}^0$ at least up to values of $K_{III}^0/K_{I}^0$ of the order of $0.15$.

\section{Linear stability analysis for arbitrary values of $K_{III}$ and constant $G_\mathrm{c}$}\label{sec:StabAnal2}

\subsection{Preliminary considerations}\label{sec:PrelimAnal2}

We briefly mentioned, at the end of Subsection \ref{subsec:GeomModeAnal1}, an issue raised by the preceding stability analysis, concerning the conditions found necessary for existence of the drifting motion of the instability modes along the crack front. This issue will now be explained and discussed in detail.

\begin{itemize}
\item Consider the case of a mixed-mode I+III loading. The absence of a drifting motion of instability modes for such a loading is rooted in symmetry properties. Indeed in the absence of mode II, both the geometry and the loading are invariant in a rotation of $180^{\circ}$ about the crack propagation direction $x$. A drifting motion of the instability mode is thus prohibited as it would violate this invariance.
\item The introduction of a mode II loading component destroys the invariance in a rotation of $180^{\circ}$ about the direction $x$, since $K_{II}^0$ changes sign in such a rotation. Hence a drifting motion of the instability modes along the crack front is no longer {\it a priori} impossible, and may reasonably be expected to occur no matter whether $G_\mathrm{c}$ depends on $\rho=K_{III}/K_I$ or not. But, surprisingly, the preceding stability analysis says otherwise since it predicts that a mode III-dependent $G_\mathrm{c}$, in addition to a nonzero $K_{II}^0$, is actually necessary for the instability modes to drift.
\end{itemize}

In the following, we revisit the preceding analysis by assuming $G_\mathrm{c}$ to be a constant, but considering arbitrary large values of the unperturbed SIF $K_{III}^0$. Our primary objective is to investigate whether or not, under such conditions, higher-order terms disregarded in the preceding analysis may lead to a drift of instability modes.


\subsection{New hypotheses}\label{sec:NewHypAnal2}

The following modified assumptions are therefore made: first, {\it the critical energy-release-rate $G_\mathrm{c}$ is assumed to be a constant}, independent of the ratio $\rho=K_{III}/K_{I}$; second, {\it the hypothesis of smallness of the ratio $\rho^0=K_{III}^0/K_{I}^0$ is relaxed}, this ratio being now allowed to take arbitrary (positive) values. Note however that the preceding hypothesis of smallness of the ratio $\varphi^0=K_{II}^0/K_{I}^0$ is retained, for the reason explained in Subsection \ref{subsec:HypGeomLoad}.

But dropping the assumption of smallness of $\rho^0$ brings back the difficulty, mentioned in Subsection \ref{subsec:HypAnal1}, that the term $\delta_y K_I^{\rm skew}(x,z)$ of the expression (\ref{eqn:Kpert})$_1$ of $K_{I}(x,z)$ has no chance of varying exponentially with the distance $x$ of propagation, if nothing more than the boundedness of $\widetilde{\phi_y}$ is assumed for this perturbation in the region $x<0$.

We therefore introduce a third assumption aimed at solving this difficulty: we consider only {\it distances $x$ of propagation of the crack much larger than the typical distance of growth $1/({\rm Re}\,\lambda)$ of the crack perturbation}. With such a hypothesis, the perturbation $\widetilde{\phi_y}$ quickly decreases behind the crack front, so that its values in the region $x<0$ have negligible impact upon that of $\delta_y K_I^{\rm skew}(x,z)$. It thus becomes harmless to use equation (\ref{eqn:ComplexInstMode})$_2$ for $\widetilde{\phi_y}$ {\it even in the region $x<0$}, and $\delta_y K_I^{\rm skew}(x,z)$ will be seen to then vary exponentially with $x$, as desired, just like all other terms in the expressions of the perturbations of the SIFs.

Clearly, this additional assumption does not permit to predict the evolution of the crack perturbation for values of the distance $x$ of propagation smaller than, or of the order of $1/({\rm Re}\,\lambda)$. It is however thought to be harmless in that the predicted evolution for $x\gg 1/({\rm Re}\,\lambda)$ becomes more and more accurate as the crack propagates.


\subsection{Application of the double criterion}\label{sec:ApplicCritAnal2}

First, one must apply \cite{GS74}'s PLS. No approximation was made in Subsection \ref{subsec:ApplicPLSAnal1} when equating the expression (\ref{eqn:K2Pert}) of $K_{II}(x,z)$ to zero, since this expression did not involve any simplification. Hence the Fourier transforms $\widehat{\psi_x}$ and $\widehat{\psi_y}$ are still tied by relation (\ref{eqn:PhiyVsPhix}).

In order to next apply \cite{G20}'s criterion, one needs the expressions of the perturbations $\delta K_{I}(x,z)$, $\delta K_{III}(x,z)$ of $K_{I}(x,z)$ and $K_{III}(x,z)$. The expression (\ref{eqn:PertK13})$_2$ of $\delta K_{III}(x,z)$ given in Subsection \ref{subsec:ApplicPLSAnal1} still applies since it did not involve any simplification either. But the expression (\ref{eqn:PertK13})$_1$ of $\delta K_{I}(x,z)$, which was obtained by discarding several terms no longer negligible here, must be corrected. The first task is to calculate the term $\delta_y K_I^{\rm skew}(x,z)$ defined by equation (\ref{eqn:dyK1skew1}) for a perturbation $\widetilde{\phi_y}(x,z)$ given by equation (\ref{eqn:ComplexInstMode})$_2$ (for all values of $x$, not only positive ones). This is done in Appendix \ref{app:DeltaK1Skew}, with the following result:
\begin{equation}\label{eqn:dK1skewFourier}
\delta_y K_I^{\rm skew}(x,z) = \frac{1}{\sqrt{2}}\, \frac{1-2\nu}{1-\nu} \, {\rm Re} \left\{ e^{\lambda x} \int_{-\infty}^{+\infty}
\left[-(1-\nu)K_{II}^0|k|+iK_{III}^0k \right]F({\lambda}/{|k|})\widehat{\psi_y}(k)e^{ikz}dk \right\}
\end{equation}
where $F$ is the function defined by
\begin{equation}\label{eqn:DefF}
F(\xi) \equiv \frac{1}{\sqrt{1+\xi}}\,,
\end{equation}
with a cut of the complex square root along the half-line of negative reals. Including other terms discarded in equation (\ref{eqn:PertK13})$_1$, we finally get for $\delta K_{I}(x,z)$:
\begin{equation}\label{eqn:PertK1}
\begin{array}{lll}
\ds \frac{\delta K_I(x,z)}{K_I^0} & = & \ds {\rm Re} \left\{ e^{\lambda x} \int_{-\infty}^{+\infty} \left[
- \frac{|k|}{2}\,\widehat{\psi_x}(k) + \left( - \frac{3}{2}\,\varphi^0\lambda -2i\rho^0k + \frac{\varphi^0}{2}\,|k| \right. \right. \right. \\[3mm]
{} & {} & \ds \left. \left. \left. + \frac{1}{\sqrt{2}}\,\frac{1-2\nu}{1-\nu}\,\left(i\rho^0k - (1-\nu)\varphi^0|k|\right)F({\lambda}/{|k|}) \right) \widehat{\psi_y}(k) \right] e^{ikz}dk \right\}.
\end{array}
\end{equation}

Since $G_\mathrm{c}$ is assumed here to be a constant, applying \cite{G20}'s criterion is equivalent to simply enforcing the condition $\delta G(x,z) \equiv 2\frac{1-\nu^2}{E}K_I^0\delta K_I(x,z) + 2\frac{1+\nu}{E}K_{III}^0\delta K_{III}(x,z) = 0$. Using equations (\ref{eqn:PertK1}), (\ref{eqn:PertK13})$_2$ and (\ref{eqn:PhiyVsPhix}), one gets from there, after a lengthy calculation, the condition
\begin{equation}\label{eqn:xiAnal2}
\xi = \frac{N_1'(\xi)+iN_2'(\xi)}{D_1'+iD_2'}
\ , \
\left\{
\begin{array}{lll}
N_1'(\xi) & \equiv & \ds -(1-\nu)(2-3\nu) + \left[ 3(2-\nu) - 4\sqrt{2}(1-2\nu)F(\xi) \right](\rho^0)^2 \\
N_2'(\xi) & \equiv & \ds \left[ 4(1-\nu) - \sqrt{2}(1-2\nu)(2-\nu)F(\xi) \right]\varphi^0\rho^0s \\
D_1' & \equiv & (1-\nu)(2-\nu) + (2+\nu)(\rho^0)^2 \\
D_2' & \equiv & \ds 8(1-\nu)\varphi^0\rho^0s
\end{array}
\right.
\end{equation}
where $s$ denotes the sign of $k$ like in equation (\ref{eqn:xiAnal1}).\footnote{The coherence of this result with that obtained in the preceding stability analysis, based on different hypotheses, may be assessed by taking $\gamma$ and $X$ nil in equation (\ref{eqn:xiAnal1}), and retaining only terms of first order in the pair $(\varphi^0,\rho^0)$ in equation (\ref{eqn:xiAnal2}); both equations then simply reduce to $\xi=-\frac{2-3\nu}{2-\nu}$.}

Unlike equation (\ref{eqn:xiAnal1}), equation (\ref{eqn:xiAnal2}) does not directly provide the value of the normalized complex growth rate $\xi$ of the perturbation, since $\xi$ also appears in the numerator $N_1'(\xi)+iN_2'(\xi)$. However the terms $3(2-\nu) - 4\sqrt{2}(1-2\nu)F(\xi)$ in the expression of $N_1'(\xi)$, and $4(1-\nu)-\sqrt{2}(1-2\nu)(2-\nu)F(\xi)$ in the expression of $N_2'(\xi)$, may be checked to vary relatively modestly with $\xi$. Hence equation (\ref{eqn:xiAnal2}) is in a form suitable for a numerical algorithm of solution based on a simple fixed-point algorithm, wherein the value $\xi^{(i)}$ of $\xi$ at iteration $i$ is obtained from that at iteration $i-1$, $\xi^{(i-1)}$, through the formula $\xi^{(i)} = \frac{N_1'(\xi^{(i-1)})+iN_2'(\xi^{(i-1)})}{D_1'+iD_2'}$, up to convergence.

Once solved numerically, equation (\ref{eqn:xiAnal2}) permits to discuss stability issues, notably the onset of instability corresponding to the vanishing of the real part of $\xi$, and the geometry of instability modes, following the same lines as in Subsection \ref{subsec:GeomModeAnal1}. With regard to the second point, the existence of imaginary parts in the numerator $N_1'(\xi)+iN_2'(\xi)$ and the denominator $D_1'+iD_2'$ of the expression of $\xi$ implies that instability modes must drift along the crack front as it propagates, just like in the preceding stability analysis. However the drifting motion no longer arises from a dependence of $G_\mathrm{c}$ upon $\rho=K_{III}/K_{I}$, assumed absent in the present analysis, but from terms proportional to $\varphi^0\rho^0=\frac{K_{II}^0K_{III}^0}{(K_{I}^0)^2}$, which were disregarded in the preceding analysis based on the hypothesis of small $\varphi^0$ and $\rho^0$.


\subsection{Numerical illustrations}\label{sec:NumExAnal2}

Again, we shall consider only values of the ratio $\varphi^0=K_{II}^0/K_{I}^0$ not exceeding $0.1$, for the reason explained in Subsection \ref{subsec:HypGeomLoad}.

\begin{figure}[h]
\centerline{\includegraphics[height=6cm]{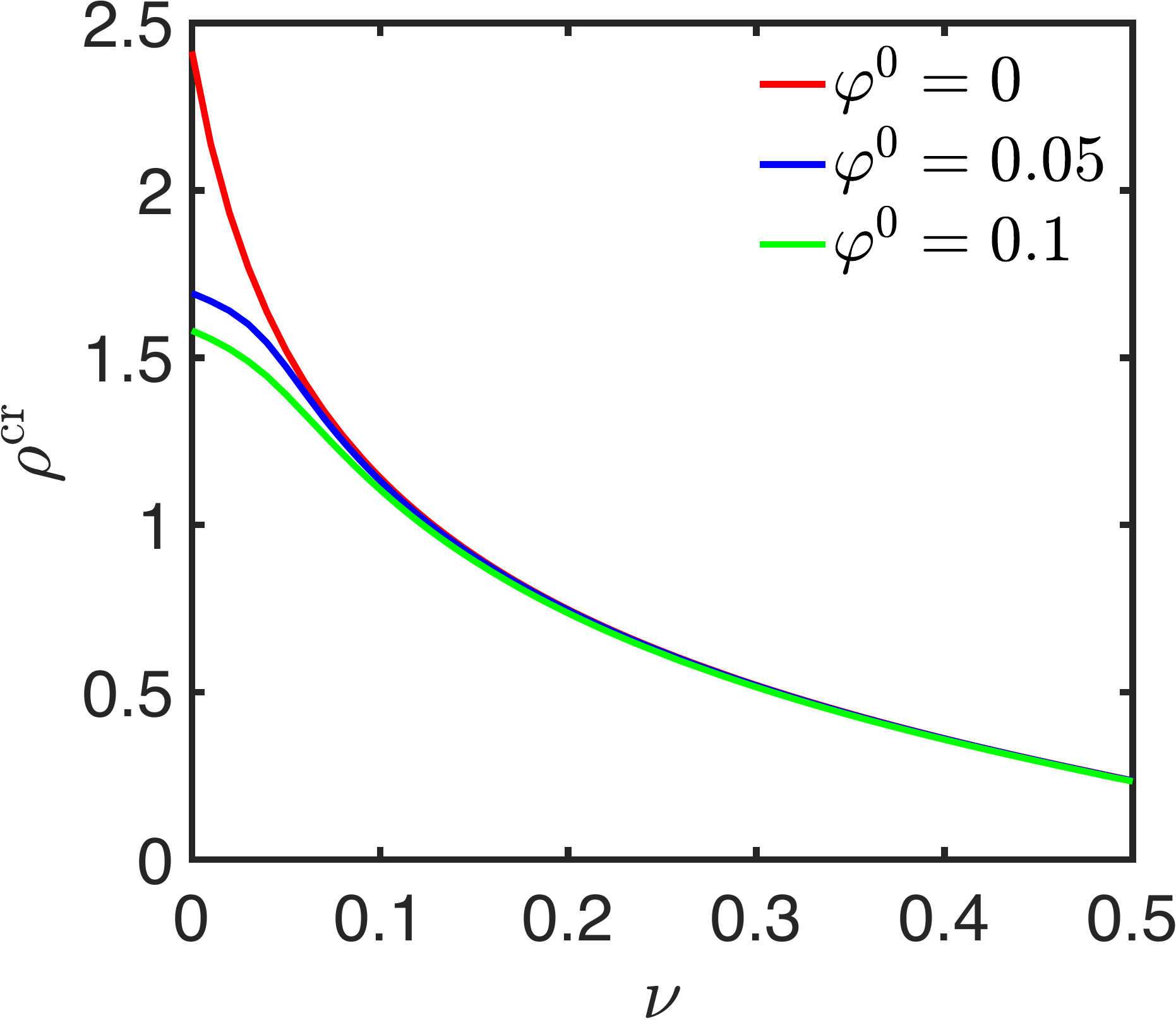}} \caption{Instability threshold $\rho^{\rm cr}$ versus Poisson's ratio $\nu$, for several values of $\varphi^0=K_{II}^0/K_{I}^0$. Here, the toughness is assumed to be constant, independent of the magnitude of the anti-plane shear component $K_{III}^0$ ($\gamma = 0$).}
\label{fig:RhocVsNuConstGc}
\end{figure}

Figure \ref{fig:RhocVsNuConstGc} first shows the threshold value $ \rho^{\mathrm{cr}}$ corresponding to incipient instability (determined from its defining condition $\xi_1={\rm Re}\,\xi=0$), versus Poisson's ratio $\nu$, for different values of the ratio $\varphi^0$. The small mode II loading component may be observed to have only a minor impact upon this threshold value, at least for usual values of Poisson's ratio exceeding $0.1$.\footnote{It is also worth noting that for $ \varphi^0 = 0$, the solution exactly coincides with the analytical one determined by \cite{LKL11}.}

\begin{figure}[h]
\centering
{
\includegraphics[height=5.5cm]{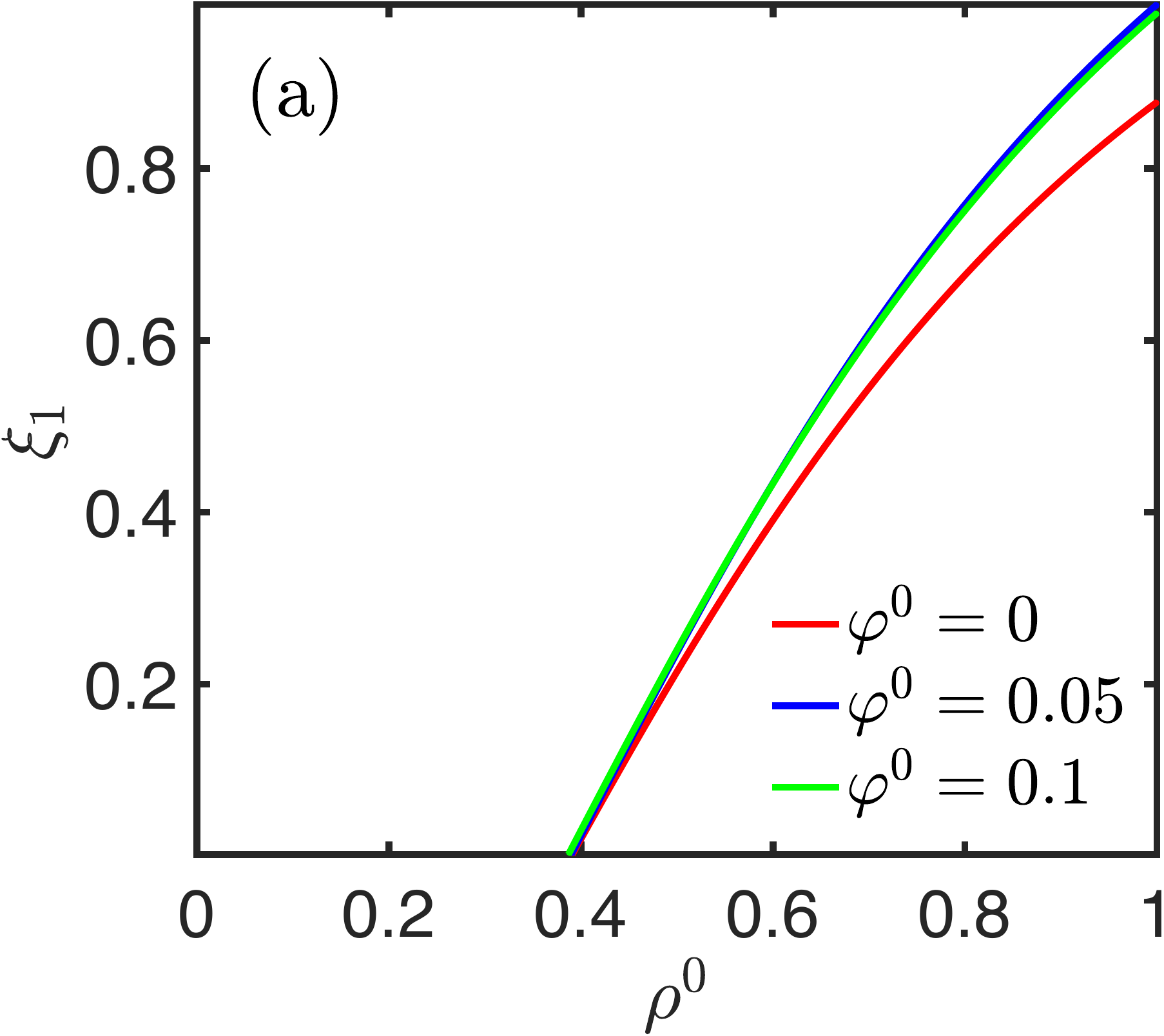}
}
{
\includegraphics[height=5.6cm]{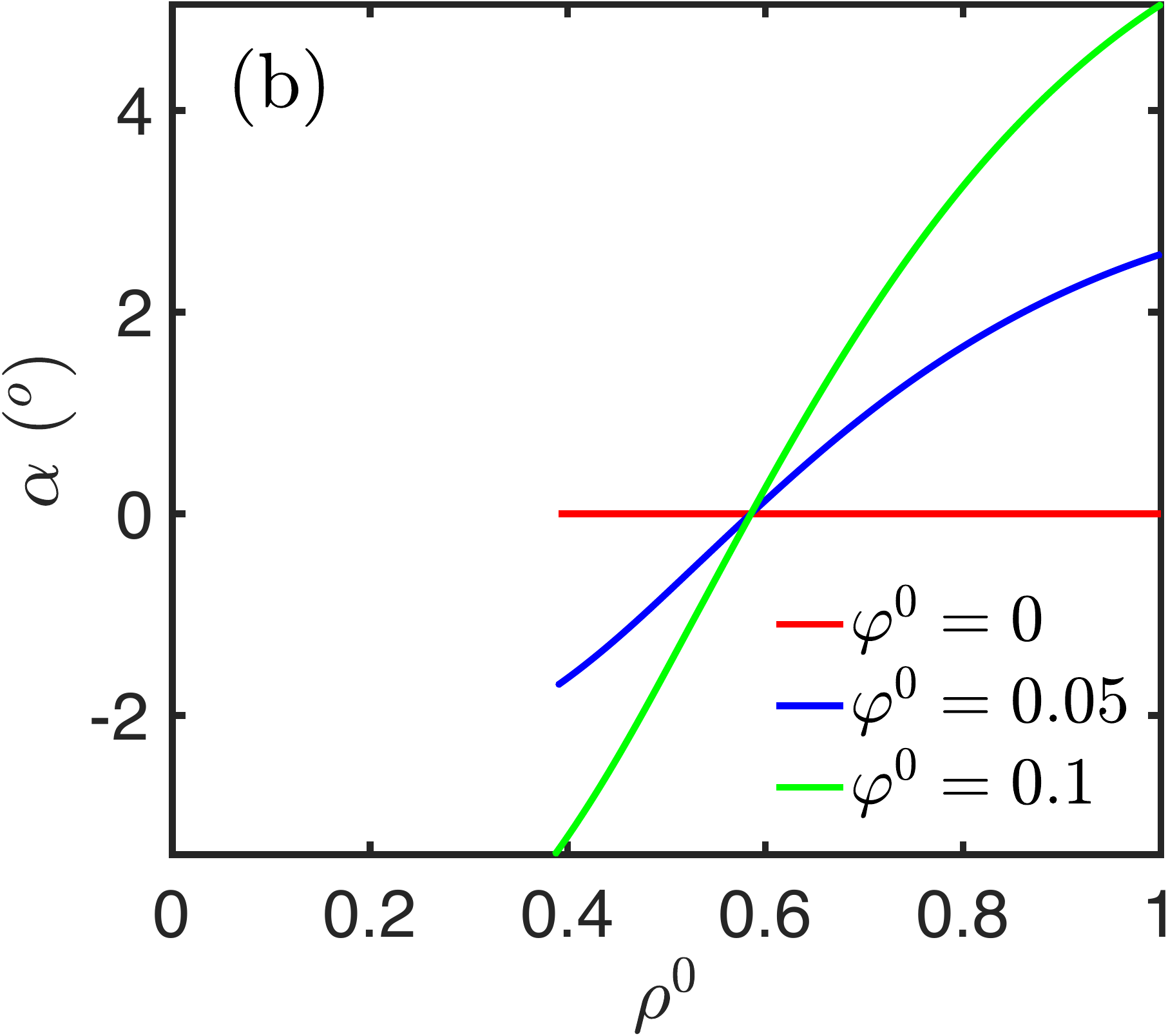}
}
\caption{\label{fig:ReImXiConstGc} Growth rate $\xi_1 = \mbox{Re } \xi$ and drift angle $\alpha = - \arctan(\mbox{Im }\xi)$ of the instability mode as functions of the ratio $\rho^0=K_{III}^0/K_{I}^0$, for several values of $\varphi^0=K_{II}^0/K_{I}^0$. }
\label{fig:DzDxConstGc}
\end{figure}

Figure~\ref{fig:ReImXiConstGc} shows the growth rate and drift angle of the instability modes as functions of the ratio $\rho^0=K_{III}^0/K_{I}^0$, for $\nu=0.38$ and various values of $\varphi^0=K_{II}^0/K_{I}^0$. Figure~\ref{fig:DzDxConstGc}(a) shows that the growth rate $\xi_1=\mbox{Re }\xi$ depends only slightly on the amount of mode II when the toughness is constant. Figure~\ref{fig:DzDxConstGc}(b) confirms that the facets do not drift in the absence of a plane shear loading component ($\varphi^0 = 0$). However, in the presence of mode II, a drifting motion is predicted, with an angle that increases (in absolute value) with $K_{II}^0$. Interestingly, the sign of the drift velocity is not only set by the sign of $K_{II}^0/K_{III}^0$, in contrast to the situation investigated in Section~\ref{sec:StabAnal1} limited to small values of $K_{III}^0$: for a positive $\varphi^0$, while $\rho^0 \gtrsim 0.58$ leads to a positive drift angle in agreement with the results of Section~\ref{sec:StabAnal1}, $\rho^0 \lesssim 0.58$ leads to a negative drift angle. This means that even in the absence of mode III-induced toughening, the direction of the drift is a subtle feature that depends upon the magnitude of the mode III loading component.\footnote{The sign of the drift angle predicted for values of $\rho^0 \gtrsim \rho^{\rm cr}$ close to the threshold should be taken with caution, since the condition $x \gg 1/(\mbox{Re }\lambda) = 1/(\xi_1 |k|)$ may not be satisfied as $\xi_1$ is then only {\it slightly} positive (see Section~\ref{sec:NewHypAnal2}).}

Finally, comparing Figs. \ref{fig:DzDx} and \ref{fig:DzDxConstGc}, one sees that instability modes drift with a much larger angle when the fracture energy depends on $\rho=K_{III}/K_I$, even for moderate values of the parameter $\gamma$ characterizing this dependence. This stems from the fact that in the analysis assuming a $\rho$-dependent $G_\mathrm{c}$, the drifting motion of instability modes is already apparent at the first order in the pair $(\varphi^0,\rho^0)$, whereas in that assuming a constant $G_\mathrm{c}$ the effect results only from higher order terms.


\section{Discussion}\label{sec:Discuss}

The application of the preceding analyses to the interpretation of experimental observations is now discussed in qualitative terms.

A preliminary remark is that fracture rarely takes place under pure tension. For instance, naturally fractured rocks observed on-site very often result from mixed-mode loading conditions (see {\it e.g.} \cite{Peacock} for a review of this feature). Nominally tensile fracture tests carried in the laboratory suffer from unavoidable misalignments of the loading system that generate small, but finite in-plane and anti-plane shear components. Fracture tests specifically designed to investigate crack propagation under mixed-mode I + III very often involve a small mode II component~\citep{LBFW08,LMR10,GO12}.

As a result, a complete picture of the fragmentation instability in mode I+III requires a detailed analysis of the effect of mode II, both on the value of the threshold $\rho^\mathrm{cr}$ and the fragmentation pattern.

Our theoretical analyses suggest that the presence of mode II affects only marginally the instability threshold as long as the material parameter $\gamma$ describing the toughening resulting from mode III is moderate, in the range $\gamma \lesssim 10$. However, when larger values of $\gamma$ are considered, the effect becomes significant. For Homalite, the material used in \cite{LMR10}, for which $\gamma \simeq 25$ can be estimated from a fit of (\ref{eqn:GcVsRho}) to experimental measurements of fracture onset under mixed-mode I+III (Fig. 6 (a) in~\citep{LMR10} and Fig. 1 in~\citep{LKPV18}), the presence of an in-plane shear component of relative amplitude $K_{II}^0/K_{I}^0 = 0.1$ results in an increase of approximatively $10~\%$ of the instability threshold. For the value $\gamma = 100$, the effect is much stronger as the same amount of mode II results in a threshold twice as large. Overall, the presence of an in-plane shear component {\it stabilizes} the crack front in the presence of anti-plane shear, and this effect must be taken into account for materials displaying a strong mode III-induced toughening (typically $\gamma \gg 10$) for the accurate prediction of the instability threshold. It is important to note here that from an applied perspective, the absence or presence of crack front fragmentation has a notable impact upon the actual material toughness and fatigue crack speed under mixed-mode loading conditions~\citep{YM89,ERK17}, so that reliably predicting the instability threshold is of major importance for the safety design and the prediction of the lifetime of structures.

But the most spectacular effect of the presence of a mode II component during crack fragmentation is on the resulting fracture pattern. Our theoretical analyses suggest that in the presence of plane shear, the crest of the helical perturbations that form above the instability threshold drifts along the front, leaving behind it ridges that are not parallel to the mean direction of crack propagation. The drift angle is set by the values of the ratios $K_{II}^0/K_{I}^0$ and $K_{III}^0/K_{I}^0$ and the material parameter $\gamma$.

The drifting of facets has been observed in various experimental studies \citep{LBFW08,BCMR08,Sherman,LMR10,Ronsin,PR14,KCF18}. It is not {\it a priori} clear, however, whether or not some or all of these observations may be ascribed to the presence of mode II during crack growth.

But a closer look at the experimental observations of~\cite{LMR10} is instructive, as they carried 3-point bending tests with a {\it controlled} mode II loading component. In these fracture tests, the mode III component was introduced by machining the notch {\it non-perpendicularly} to the sample surface (see Fig. \ref{fig:ExpSetup} (a)). This resulted in a mode II component that varied linearly along the front (Fig. \ref{fig:ExpSetup} (b)), while the mode III component remained essentially uniform along it and positive. The sign of $K_{II}^0$ is not constant and is positive in the region $z>0$ while negative in the region $z<0$. The resulting fracture pattern was remarkable (see Fig. \ref{fig:ExpSetup}(c)): the facets drifted towards opposite directions in the two halves of the specimens, with a positive drift angle $\alpha = \arctan(dz/dx)$ in the region $z>0$ where the signs of $K_{II}^0$ and $K_{III}^0$ were identical, and a negative one in the region $z<0$ where these signs were opposite. These observations are in qualitative agreement with the predictions of our theoretical stability analyses.

To go beyond these qualitative observations, one must examine the magnitude of the drift. Characteristic relative values of the shear loading components imposed upon the specimen during fracture, at a point of the crack front located at some distance from the mid-plane $z=0$, were $K_{II}^0/K_{I}^0 \simeq 0.08$ and $K_{III}^0/K_{I}^0 \simeq 0.08$, leading to $K_{II}^0/K_{III}^0 \simeq 1$. For these parameters and a value of Poisson's ratio of $0.34$ (typical of Homalite), equation~\eqref{eqn:xiAnal2}) for a constant fracture energy predicts a drift angle $\alpha \simeq 2^\circ$ which is way too small, while equation \ref{eqn:LaterVeloc} for a mode III-dependent fracture energy predicts, with the value $\gamma \simeq 25$ determined experimentally \citep{LMR10,LKPV18}, $\alpha \simeq 16^\circ$ which is still smaller than, but nevertheless much more compatible with, the experimental drift angle $\alpha \simeq 20 - 30^\circ$ (see Fig. \ref{fig:ExpSetup}(d)).

Obviously, much more numerous and detailed comparisons between experimental and predicted values of the drift angle (currently under progress) are necessary to more comprehensively assess the validity of the theory developed.

\section{Conclusion}\label{sec:Concl}

The configurational stability of a crack propagating under fully general mixed-mode (I+II+III) loading conditions was investigated within the classical framework of LEFM, on the basis of a linear stability analysis. This analysis stood as a natural extension of that of~\cite{LKPV18} that was limited to mode I+III. A mode III-dependent fracture energy was postulated like in the previous work, and its effect on both the fragmentation threshold and the geometry of bifurcated modes was analyzed.

The main findings of the new study, and the perspectives it opens, are as follows:
\begin{itemize}
\item The presence of mode II {\it stabilizes} the crack front, as the fragmentation threshold $\rho^\mathrm{cr}$ (critical value of the ratio $\rho^0 = {K_{III}^0}/{K_{I}^0}$ of the unperturbed mode III to mode I SIFs) increases with the unperturbed SIF of mode II, $K_{II}^0$, whether the fracture energy is mode-dependent or not. This effect becomes significant for large values of both the parameter $\gamma$ characterizing the shear-induced toughening {\it and} the ratio $\varphi^0 = {K_{II}^0}/{K_{I}^0}$ of the unperturbed mode II to mode I SIFs.
\item The presence of mode II has a strong impact upon the geometry of the crack front in the unstable regime. In the presence of in-plane shear, the tops of the helical perturbations that form above the instability threshold drift along the front as the crack propagates, leaving behind them ridges oriented at an angle with the mean direction of crack propagation.
\item The value of the drift angle thus defined depends in a non-trivial way upon those of the ratios $K_{II}^0/K_{I}^0$ and $K_{III}^0/K_{I}^0$ and the material parameter $\gamma$.
\item The drift angle is predicted to strongly increase with the parameter $\gamma$, and be much larger when the fracture energy is mode-dependent ($\gamma > 0$) than when it is not ($\gamma = 0$).
\item The comparison of the geometrical features of the fragmentation pattern predicted theoretically and actually observed offers interesting perspectives, which will be fully pursued in some future work.
\item Another perspective would consist of performing new numerical simulations based on \cite{KKL01}'s phase-field model, now under fully general I+II+III mixed-mode conditions. The toughening of the material induced by the presence of mode III could be included through some suitable extension of the model. The aim of such simulations would be to assess the importance of geometrical nonlinearities disregarded in the present paper.
\end{itemize}

{\bf Acknowledgements}

L.P. and A.V. acknowledge the support of the City of Paris through the Emergence Program.
A.K. acknowledges
support of Grant DEFG02-07ER46400 from the US Department
of Energy, Office of Basic Energy Sciences.


\appendix

\section{Appendix - Calculation of $\delta_y K_I^{\rm skew}$ for a perturbation of the form (\ref{eqn:ComplexInstMode})$_2$}\label{app:DeltaK1Skew}

To calculate $\delta_y K_I^{\rm skew}(x,z)$ for a perturbation $\widetilde{\phi_y}(x',z')$ of the form (\ref{eqn:ComplexInstMode})$_2$ (everywhere in the complex plane, not only in the region $x'>0$), the first task is to evaluate it for the complex perturbation $e^{\lambda x'}\,\widehat{\psi_y}(k)e^{ikz'}$.\footnote{Equation (\ref{eqn:dyK1skew1}) {\it a priori} defines $\delta_y K_I^{\rm skew}(x,z)$ only for real perturbations, but is readily formally extended to complex ones - preserving linearity - by simply retaining the same formula.} For this perturbation equation (\ref{eqn:dyK1skew1}) yields
\begin{equation*}
  \begin{array}{l}
    [\delta_y K_I^{\rm skew}(x,z)]_{|\,\widetilde{\phi_y}(x',z')=e^{\lambda x'}\,\widehat{\psi_y}(k)e^{ikz'}} = \\[3mm]
    \ds \frac{\sqrt{2}}{4\pi} \, \frac{1-2\nu}{1-\nu} \, {\rm Re} \left\{ \int_{-\infty}^{x} dx' \int_{-\infty}^{+\infty} \frac{ \left[ K_{III}^0-i(1-\nu)K_{II}^0 \right] e^{\lambda x'} \widehat{\psi_y}(k).\,ike^{ikz'}}{(x-x')^{1/2}\left[x-x'+i(z-z')\right]^{3/2}} \, dz' \right\} = \\[3mm]
    \ds \frac{\sqrt{2}}{4\pi} \, \frac{1-2\nu}{1-\nu}\, {\rm Re} \left\{ i\left[ K_{III}^0-i(1-\nu)K_{II}^0 \right] e^{\lambda x} k\widehat{\psi_y}(k) \int_{0}^{+\infty} \frac{e^{-\lambda v}}{\sqrt{v}} \,dv \int_{-\infty}^{+\infty} \frac{e^{ikz'}}{\left[v+i(z-z')\right]^{3/2}} \, dz' \right\}
  \end{array}
\end{equation*}
where the change of variable $v\equiv x-x'$ has been used. The integral over $z'$ here has been evaluated by \cite{LKL11} using complex analysis, and the result is:
\begin{equation*}
  \int_{-\infty}^{+\infty} \frac{e^{ikz'}}{\left[v+i(z-z')\right]^{3/2}} \, dz' = 4H(-k)\sqrt{\pi |k|}\,e^{-|k|v}e^{ikz}
\end{equation*}
where $H$ denotes Heaviside's function. Therefore
\begin{equation*}
  \begin{array}{l}
    \ds [\delta_y K_I^{\rm skew}(x,z)]_{|\,\widetilde{\phi_y}(x',z')=e^{\lambda x'}\,\widehat{\psi_y}(k)e^{ikz'}} = \\[3mm]
    \ds \sqrt{\frac{2}{\pi}} \, \frac{1-2\nu}{1-\nu}\, H(-k)k\sqrt{|k|}\,{\rm Re} \left\{i\left[ K_{III}^0-i(1-\nu)K_{II}^0 \right] e^{\lambda x} \widehat{\psi_y}(k)e^{ikz} \int_{0}^{+\infty} \frac{e^{-(|k|+\lambda)v}}{\sqrt{v}} \,dv \right\}.
  \end{array}
\end{equation*}
For {\it real} $\lambda$, the integral on $v$ here is readily calculated using the change of variable $w \equiv \sqrt{(|k|+\lambda)v}$, with the following result:
\begin{equation*}
  \int_{0}^{+\infty} \frac{e^{-(|k|+\lambda)v}}{\sqrt{v}} \,dv = \sqrt{\frac{\pi}{|k|}} \, F(\lambda/|k|) \quad {\rm where} \quad F(\xi)\equiv\frac{1}{\sqrt{1+\xi}} \,.
\end{equation*}
But both the left- and right-hand-sides here are analytic functions of the parameter $\lambda$ spanning the whole complex plane except the half-line of real numbers $\leq -|k|$. Therefore the equality in fact holds over the domain thus defined, the cut of the complex square root being along the half-line of negative reals. It follows that
\begin{equation*}
  \begin{array}{l}
    \ds [\delta_y K_I^{\rm skew}(x,z)]_{|\,\widetilde{\phi_y}(x',z')=e^{\lambda x'}\,\widehat{\psi_y}(k)e^{ikz'}} = \\[3mm]
    \ds \sqrt{2}\, \frac{1-2\nu}{1-\nu} \, H(-k)k\,{\rm Re} \left\{i\left[ K_{III}^0-i(1-\nu)K_{II}^0 \right] e^{\lambda x} F(\lambda/|k|) \widehat{\psi_y}(k)e^{ikz} \right\}.
  \end{array}
\end{equation*}

For the conjugate perturbation $e^{\overline{\lambda} x'}\,\overline{\widehat{\psi_y}(k)}\,e^{-ikz'}$, one similarly gets (since $F(\,\overline{\lambda}/|k|)=\overline{F(\lambda/|k|)}$ ):
\begin{equation*}
  \begin{array}{l}
    \ds [\delta_y K_I^{\rm skew}(x,z)]_{|\,\widetilde{\phi_y}(x',z')=e^{\overline{\lambda} x'}\,\overline{\widehat{\psi_y}(k)}\,e^{-ikz'}} = \\[3mm]
    \ds \sqrt{2}\, \frac{1-2\nu}{1-\nu} \, H(k)(-k)\,{\rm Re} \left\{i\left[ K_{III}^0-i(1-\nu)K_{II}^0 \right] e^{\overline{\lambda} x} \overline{F(\lambda/|k|)} \ \overline{\widehat{\psi_y}(k)}\,e^{-ikz} \right\} = \\[3mm]
    \ds \sqrt{2}\, \frac{1-2\nu}{1-\nu} \, H(k)k\,{\rm Re} \left\{i\left[ K_{III}^0+i(1-\nu)K_{II}^0 \right] e^{\lambda x} F(\lambda/|k|)\widehat{\psi_y}(k)e^{ikz} \right\}.
  \end{array}
\end{equation*}
It then follows from linearity that for the perturbation ${\rm Re}[e^{\lambda x'}\,\widehat{\psi_y}(k)e^{ikz'}]$,
\begin{equation*}
  \begin{array}{l}
    \ds [\delta_y K_I^{\rm skew}(x,z)]_{|\,\widetilde{\phi_y}(x',z')={\rm Re}[e^{\lambda x'}\,\widehat{\psi_y}(k)e^{ikz'}]} = \\[3mm]
    \ds \frac{1}{2}\,\left\{ [\delta_y K_I^{\rm skew}(x,z)]_{|\,\widetilde{\phi_y}(x',z')=e^{\lambda x'}\,\widehat{\psi_y}(k)e^{ikz'}}
    + [\delta_y K_I^{\rm skew}(x,z)]_{|\,\widetilde{\phi_y}(x',z')=e^{\overline{\lambda} x'}\,\overline{\widehat{\psi_y}(k)}\,e^{-ikz'}} \right\} = \\[3mm]
    \ds \frac{1}{\sqrt{2}}\, \frac{1-2\nu}{1-\nu} \,{\rm Re} \left\{\left[-(1-\nu)K_{II}^0|k|+iK_{III}^0k\right] e^{\lambda x} F(\lambda/|k|) \widehat{\psi_y}(k)e^{ikz} \right\}
  \end{array}
\end{equation*}
where the relations $H(k)+H(-k)=1$, $H(k)-H(-k)={\rm sgn}(k)$ have been used.

It finally follows from this relation and linearity that for the perturbation $\widetilde{\phi_y}$ given by equation (\ref{eqn:ComplexInstMode})$_2$ (everywhere in the complex plane),
\begin{equation*}
    \delta_y K_I^{\rm skew}(x,z) = \frac{1}{\sqrt{2}}\, \frac{1-2\nu}{1-\nu} \,{\rm Re} \left\{ e^{\lambda x} \int_{-\infty}^{+\infty} \left[ -(1-\nu)K_{II}^0|k|+iK_{III}^0k \right] F(\lambda/|k|) \widehat{\psi_y}(k)e^{ikz}dk \right\}
\end{equation*}
which is equation (\ref{eqn:dK1skewFourier}) of the text. 

\end{document}